\documentclass[aps,preprint,showpacs,superscriptaddress,groupedaddress]{revtex4}  %
\usepackage{graphicx}  %
\usepackage{dcolumn}   %
\usepackage{bm}        %
\usepackage{amssymb}   %
\usepackage{nicefrac}
\usepackage{mathrsfs}
\usepackage{slashed} 
\usepackage{array} 
\usepackage{epsfig}
\usepackage{cmap} 

\newcommand{\bra}[1]{\big\langle \, #1\,\big\vert}
\newcommand{\ket}[1]{\big\vert\, #1\,\big\rangle}
\newcommand{\bracket}[2]{\big\langle \, #1 \big\vert \, #2\,\big\rangle}

\newcommand{\tg}{\mathop{\mathrm{tg}}}
\newcommand{\ch}{\mathop{\mathrm{ch}}}
\newcommand{\sh}{\mathop{\mathrm{sh}}}

\begin{document}

\title{Test of a hypothesis of realism in quantum theory using a bayesian approach}

\author{N. Nikitin}
\affiliation{Lomonosov Moscow State University Department of Physics, Russia}
\affiliation{Lomonosov Moscow State University Skobeltsyn Institute of
  Nuclear Physics, Russia}
\affiliation{Institute for Theoretical and Experimental Physics,
  Russia}
\affiliation{National Research Nuclear University MEPhI, Russia}
\author{K. Toms}
\affiliation{Department of Physics and Astronomy, University of New
  Mexico, USA}
\date{\today}

\begin{abstract}
In this paper we propose a time-independent \textit{equality} and
time-dependent \textit{inequality}, suitable for an experimental test
of the hypothesis of realism. The derivation of these relations is based on 
the concept of conditional probability and on Bayes' theorem in the
framework of Kolmogorov's axiomatics of probability theory. The 
equality obtained is intrinsically different from the well known GHZ-equality
and its variants, because violation of the new equality might be
tested in experiments with only two microsystems in a maximally
entangled Bell state $\ket{\Psi^-}$, while a test of the GHZ-equality
requires at least three quantum systems in a special state
$\ket{\Psi^{\textrm{GHZ}}}$. The obtained inequality differs from
Bell's, Wigner's, and Leggett-Garg inequalities, because it deals with spin
$s=1/2$  projections onto only two non-parallel directions at two
different moments of time, while a test of the Bell and Wigner
inequalities requires at least three non-parallel directions, and a
test of the Leggett-Garg inequalities -- at least three distinct  moments
of time. Hence, the proposed inequality seems to allow one to avoid the
``contextuality loophole''. Violation of the proposed equality and
inequality is illustrated with the behaviour of a pair of
anti-correlated spins in an external magnetic field and also with the
oscillations of flavour-entangled pairs of neutral pseudoscalar
mesons. 
\end{abstract}

\maketitle

\section{Introduction}

One of the main questions that arise when comparing predictions
of quantum theory with experiment is to what extent the real physical
properties of micro-objects correspond to the observed values, measured
with macro-devices. Heisenberg noted in his book
\cite{heisenberg1958}  that this essential question of quantum theory
is close to the general analysis of our perception of ``phenomenon'' of
our world and the gist of the phenomenon, the ``noumenon'',
according to Kant \cite{kant_i}.

From the point of view of the simplest ``orthodox'' version of the quantum
mechanical formalism \cite{Dirac-1999}, a process of measurement
corresponds to expansion of a microsystem state vector $\ket{\psi}$
into a superposition of a macroscopically definitive states  $\ket{a_\alpha}$:

\begin{eqnarray}
\label{superposition}
\ket{\psi} = \sum\limits_\alpha C_\alpha \ket{a_\alpha},
\end{eqnarray}
where, according to the law of Born, $|C_\alpha |^2$
defines the probability to find the system in state $\ket{a_\alpha}$ after
measurement. Usually one takes $\ket{a_\alpha}$ as a set of
eigenvectors of hermitian operator $\hat A$ which corresponds to some
physical characteristics (observables) $A$ of the microsystem
studied. Of course, the expansion (\ref{superposition}) and
  law of Born may be generalized in terms of POVMs
(positive operator-valued measures -- a description of a measurement
using positively-defined operators) and the projection postulate of
Dirac--von~Neumann \cite{Peres-1993}. However it is not important for
the consequent arguments which approach is used. We will use the
simplest one, i.e. the superposition principle (\ref{superposition}) and
the law of Born.

Let a microsystem now have two distinct observables $A$ and $B$ which
have spectra $\{ a_\alpha \}$ and $\{ b_\beta \}$ accordingly. If
physical characteristics $A$ and $B$ may be simultaneously measured
(i.e. may be measured with zero dispersion with a pair of macroscopic
devices of the same type), then the vectors by which the state $\ket{\psi}$
is expanded must be the common eigenvectors of the operators $\hat A$
and $\hat B$, leading to the commutation condition $\left [ \hat A, \hat 
  B\right ] =0$. If the operators $\hat A$ and $\hat B$ do not
commute, then they do not have a common system of eigenvectors. In this
case the observables $A$ and $B$ can not be measured together by any
macro-device. The simplest example of the observables that can not be
measured together is the projection of a fermion spin onto two
non-parallel directions, which are defined by unit vectors $\vec a$
and $\vec b$. Another example is the $CP$-parity and flavour of a neutral
pseudoscalar meson.

The following question may be posed: do the physical characteristics
$A$ and $B$ exist simultaneously and independently without the
assumption of the possibility to measure them by some macro-devices
(this is the hypothesis of local realism). Usually the terms ``hypothesis
  of local realism'' and ``concept of macroscopic realism'' are
  understood as the possibility to describe a physical system in
  the classical paradigm using some assumptions about the nature of
  ``classical reality''. It might be for example locality or the
  negligible influence of a measurement device. All these assumptuons
  we will call together the ``hypothesis of realism'' by Einstein
\cite{Einstein:1935rr}. It does not make sense to talk about the
physical properties of a micro-object without making a
statement about the macro-devices used to measure these properties
(this is the Copenhagen interpretation of quantum mechanics and the principle of
complementarity of Bohr \cite{Bohr:1935af}).

A natural (but probably not unique) way to write in mathematical
terms the condition that a set of physical characteristics of a
micro-system exists jointly regardless of the possibility to measure it
with a macro-device, is that the joint probability of the set
of observables under consideration is non-negative at any time. For example for the
observables $A$ and $B$ that means that for any elements of the spectra
$a_{\alpha'} \in \{ a_\alpha \}$ and $b_{\beta'} \in \{ b_\beta \}$ 
the probability of simultaneous existence of $a_{\alpha'}$ and
$b_{\beta'}$ -- the joint probability $w(a_{\alpha'} \cap b_{\beta'})$
-- satisfies the following condition:
\begin{eqnarray}
\label{w(ab)}
0\,\le\, w(a_{\alpha'} \cap b_{\beta'})\,\le\, 1.
\end{eqnarray}
The assumption of the existence of non-negative joint probabilities
(\ref{w(ab)}) was implicitly used by Bell in his pioneering works
\cite{Bell:1964kc,Bell:1964fg}, as the density distribution $\rho
(\lambda)$ of hidden variables $\lambda$ is a direct corollary of
(\ref{w(ab)}). Later Bell's idea was developed by 
Clauser, Horne, Shimony, and Holt \cite{Clauser:1969ny}. 
A historical review of Bell's inequalities may be found in
\cite{epr-books,Reid:2009zz,Rosset:2014tsa,RevModPhys.86.419,QuantumUnSpeakables-II}, 
The idea of the existence of non-negative joint probabilities (\ref{w(ab)})
was used by Wigner in \cite{wigner}. In \cite{DEMUYNCK198665} the
arguments of Bell were translated into non-negative joint
probabilities for the first time.

In classical physics the joint probabilities 
$
 w(a_{\alpha'} \cap b_{\beta'}) =  w(b_{\beta'} \cap a_{\alpha'})
$
always exist and are well defined for any physical system. In quantum
theory if 
$
\left [ \hat A, \hat B\right ] \ne 0
$,
the joint probabilities $w(a_{\alpha'} \cap b_{\beta'})$ can not be
direcly measured by macro-devices. However in this case it is possible
to use an indirect procedure based on specific properties of entangled
states and the notion of an ``element of physical reality'' introduced by
Einstein.

The ``element of physical reality'' is defined as follows
\cite{Einstein:1935rr}: \textit{``If, without in any way disturbing a
  system, we can predict with certainty (i.e., with probability equal
  to unity) the value of a physical quantity, then there exists an
  element of physical reality corresponding lo this physical
  quantity.''} and \textit{``every element of the physical reality
  must have a counterpart in the physical theory''}.
It is obvious that a given element may be identified with
some property of a physical system (for example with a spin projection onto
some direction) and that obtaining information about the element of
physical reality differs from obtaining information about some
observable only by measurement method. In the first case the
measurement is indirect, in the second, direct, and
accompanied  by a reduction of the state vector or density matrix. Because
of the above we will not make any distinction in the current paper
between the observables and the elements of physical reality.

Let us show how the indirect procedure works using the decay of a
pseudoscalar meson to a fermion-antifermion pair. If the decay happens
at time $t_0$ through the strong or electromagnetic interaction
(i.e. preserving $P$-parity), then the pair will be in a spin-singlet
Bell state $\ket{\Psi^-}$. This fact follows from the general
structure of Hamiltonians 
\begin{eqnarray}
\label{Heff_for_PS2ff}
 \mathcal{H}^{(PS)}(x) & =& g\,\varphi (x)\,\left (\bar f(x)\,\gamma^5\, f(x)\right )_N, \\
\mathcal{H}^{(A)}(x)   & =& g' \left (\partial_{\mu}\varphi (x)\right ) \, \left (\bar f(x)\,\gamma^{\mu}\gamma^5\, f(x)\right )_N, \nonumber
\end{eqnarray}
which can be compared with similar decays in quantum field theory
(QFT). Here $\varphi (x)$ is the field of pseudoscalar particles, $\bar
f(x)$ and $f(x)$ are fermionic fields, $\partial_{\mu} = \partial
/ \partial x^{\mu}$ is divergence, $g$ and $g'$ are effective coupling
constants. Let us denote the antifermion with index ``1'' and the
fermion with index ``2''. Let the spin projections of the fermion and
the antifermion onto two directions exist simultaneously or
jointly. Note that the  directions in space are defined by
non-parallel unit vectors ${\vec a}$ and ${\vec b}$, such that the spin
projection operators do not commute. For brevity let us denote
the spin $1/2$ projection of fermion $i$ onto any axis, specified by
unit vector ${\vec n}$, as 
$$
s_{{\vec n}}^{(i)}\, =\,\pm\,\frac{1}{2}\,\equiv\, n_{\pm}^{(i)},
$$
where $i=\{1, 2\}$. Then the spin projections at the initial time
$t_0$ onto each of the directions in the state $\ket{\Psi^-}$ satisfy
the anticorrelation condition 
\begin{eqnarray}
\label{pm=mp2}
n_{\pm}^{(2)}(t_0)\, =\, -\, n_{\mp}^{(1)}(t_0).
\end{eqnarray}
Let us denote the spin projection operators of the fermion and
antifermion onto direction ${\vec a}$
as $\hat A^{(2)}$ and $\hat A^{(1)}$ accordingly. Similarly $\hat
B^{(2)}$ and $\hat B^{(1)}$  are the spin projection operators onto
direction ${\vec b}$. As the vectors ${\vec a}$ and ${\vec b}$ are
non-parallel 
\begin{eqnarray}
\label{No-commuting}
\left [ \hat A^{(1)}, \hat B^{(1)}\right ] \ne 0 , \quad \left [ \hat A^{(2)}, \hat B^{(2)}\right ] \ne 0. 
\end{eqnarray}
At the same time, according to Eberhard's theorem \cite{eberhard},
\begin{eqnarray}
\label{NS-conditions}
\left [ \hat A^{(1)}, \hat B^{(2)}\right ] = 0 , \quad \left [ \hat A^{(2)}, \hat B^{(1)}\right ] = 0. 
\end{eqnarray}
Equalities (\ref{NS-conditions}) ensure locality of the quantum theory
(even non-relativistic) at the level of macro-devices (so called ``non-signaling conditions'').

The commutation conditions (\ref{NS-conditions}) allow joint
measurement for example of the projection of the fermion spin onto direction
${\vec a}$ and the projection of the antifermion spin onto direction
${\vec b}$. Hence the joint probability   
$
w \left (a_\alpha^{(2)},\, b_\beta^{(1)},\, t \right )
$
at any time is a well-defined value and it is possible to use for it 
probability theory based on Kolmogorov's axiomatics. Here $\{\alpha,
\beta, \gamma\} = \{+,\, -\}$. Let us apply to
this probability the concept of the elements of physics reality and
the anticorrelation condition (\ref{pm=mp2}). Then for the time $t_0$ it
is possible to formally introduce the joint probability
\begin{eqnarray}
\label{w(ab)PhysReality}
w \left (a_\alpha^{(2)},\, b_\beta^{(2)},\, t_0 \right ) \equiv w \left (a_\alpha^{(2)},\, -\, b_{-\,\beta}^{(1)},\, t_0 \right )
\end{eqnarray}
of the existence of physical characteristics of a microsystem (in our case
-- the projection of fermionic spins onto two non-parallel directions
$\vec a$ and $\vec b$), corresponding to simultaneously non-measurable
observables $A^{(2)}$ and $B^{(2)}$. So, despite condition
(\ref{No-commuting}), the definition of the element of physical
reality allows us to give operational meaning to the joint probability $w
\left (a_\alpha^{(2)},\, b_\beta^{(2)},\, t_0 \right )$ and its
analogs. I.e., formula (\ref{w(ab)PhysReality}) might be considered
as a possible expansion of the definition of the joint probability concept
to the area where Heisenberg's uncertainty principle prevents us
from defining such a probability for direct measurements. It
seems logical to assume that Bayes' theorem can be applied to
probabilities like (\ref{w(ab)PhysReality}).

Using the concept of local realism, it is possible to derive not
only Bell- or Wigner-like inequalities, but also equalities. 
Such equalities, often called Greenberger-–Horne–-Zeilinger
(GHZ) equalities, were introduced in \cite{GHZ1989}. 
Proofs of GHZ equalities may be found in \cite{GHSZ1990,
  mermin1990}. Using the concept of local realism and some 
additional assumptions, a system of equations has been obtained for
distinct spin projections of three fermions in the
GHZ-state 
$
\ket{\Psi^{\textrm{GHZ}}}\, =\, \frac{1}{\sqrt{2}}\, \left ( \ket{n_+^{(1)}\, n_+^{(2)}\, n_+^{(3)}} - \ket{n_-^{(1)}\, n_-^{(2)}\, n_-^{(3)}} \right )
$.
This system is incompatible with calculations in
non-relativistic quantum mechanics.

Beginning in 1972 \cite{Freedman1972} there is
much experimental evidence of violation of Bell's and Wigner's
inequalities, i.e. evidence of unsoundness of the hypothesis of local
realism and/or the concept of elements of physical reality. However 
until recently these experiments were not free
from some loopholes, which cast doubt on the connection between
the violations of Bell's or Wigner's inequalities and the soundness of
the quantum mechanical description of the world. The first two are the
locality loophole (or communication loophole) and
fair-sampling loophole (or detection loophole).

Let us consider the first loophole using the example with
fermion-antifermion pairs. In this case the locality loophole appears
due to the fact that during the measurement of the spin projections of
each of the particles, the spacetime interval between them is
time-like. Hence in the process of measurement it is not possible to
exclude a hypothetical exchange of information between macro-devices,
which enables a quasi-non-local strong correlation between the two
measurements. Such a correlation may lead to the violation of Bell's
or Wigner's inequalities. The ``locality loophole''
was first overcome in experiments by Aspect
\cite{aspect1982} using the idea by Wheeler  
\cite{wheeler} of delayed choice for a pair of two-channel polarizers. 
In quantum optics this loophole is closed by ``brute force''
\cite{gisin1998}, when pairs of correlated photons are separated by a
significant distance. For example in photon experiments with fiber
cables this distance is greater than a hundred kilometers
\cite{143}. For spin-correlated fermions this distance is about 1.3
kilometers \cite{Nature526-spin-e-2015}. When analyzing experiments
\cite{143, Nature526-spin-e-2015} one should by definition take into
account the limit of the speed of signal exchange between
spatially separated subsystems of a correlated quantum system. Hence one
should use QFT where the signal exchange speed is finite,
instead of non-relativistic quantum mechanics where this speed is
formally infinite \cite{Nikitin:2009sr}.

The second loophole arises due to the fact that all the detectors have
certain registration efficiencies, and hence there is a gedanken
possibility to select from the whole
set the pairs of correlated particles which lead
to the violation of Bell's or Wigner's inequalities, ignoring the
others. For exclusion of the
detection loophole in Bell's inequalities, one needs to have
efficiency of the detector of the level 2/3 \cite{eberhard1993}. In
quantum optics this barrier was overcome only in 2013
\cite{Nature497-2013,detloophole2013}. For Wigner's inequalities the
authors are not aware of any papers studying that efficiency
value. The detection loophole should play an
important role in particle physics, because typical efficiencies of 
detectors (taking into account selection criteria) do not exceed
percents.

Up to now, in each of the experiments for testing Bell's
inequalities, it was possible to close only one of the two loopholes --
either locality or detection (\cite{RevModPhys.86.419} and
\cite{aspect2015}). However in 2015 three successful experiments were conducted
(two with photon pairs \cite{PRL-2015-250401, PRL-2015-250402}, and
one with correlated spins $s=1/2$ \cite{Nature526-spin-e-2015}, which
managed to avoid both loopholes.

Let us consider another few loopholes that have not been closed by
contemporary experiments. Also, as will be shown below, the
experimental situation chosen to avoid the locality loophole
\cite{143,Nature526-spin-e-2015,PRL-2015-250401, PRL-2015-250402},
raises questions about applying these experiments to
time-independent Bell's inequalities
\cite{Bell:1964kc,Bell:1964fg,Clauser:1969ny} and static Wigner's
inequalities \cite{wigner}.

Among the rest of the loopholes, the most quoted in the literature the
freedom-of-choice loophole \cite{bell2004}, where due to hidden
interactions and unknown parameters an experiment itself causes the
observer to select events
with stronger correlations. In our opinion such a loophole is
unfalsifiable and hence should not be
considered scientifically according to Popper's refutability
criterion \cite{popper1983}.

We now consider the ``contextuality loophole''
\cite{baere1984-1,baere1984-2}, which we believe is as
important as the two previously discussed. Let us define ``context''
as the aggregate of all the experimental conditions. Then, in order to
obtain the values of correlators in Bell's inequalities or joint
probabilities in Wigner's inequalities, it is necessary to conduct
four or three independent experiments. There is no
guarantee that the measurements are conducted under the same
conditions. Even if one uses the same experimental devices, it is
still impossible to repeat exactly all the internal parameters of
macro-devices, because, for example, in each of experiments one should
select different pairs of spin directions. I.e. it is impossible to
make sure that quantum ensembles of correlated particles are identical
in all the experiments. All the more, if the experiments were
conducted in different and non-controllable conditions, their results
should not be summed up, subtracted, or compared with each other. If
we suppose that every experiment has its own distribution of
probabilities of the observable spectra, then it is possible to
obtain a generalized Bell's inequality which is never violated in
experiment \cite{khrennikov2009}.

To avoid the contextuality loophole it is
necessary to conduct the measurements with an invariable state of
the macro-devices, i.e. with only two non-parallel directions ${\vec a}$
and ${\vec b}$. One can use time as an additional degree of
freedom. The same inequality, which potentially allows the experimenters
to avoid the contextuality loophole, will be introduced in Section
~\ref{section:mainBayes} of this paper.

Time-dependent
generalizations of static Bell's and Wigner's inequalities may be
justified in another way. The derivation of these inequalities strongly
depends not only on the simultaneous existence of all of the physical
characteristics of a micro-system, but also on the assumption of locality
both on the macroscopic measurement level (i.e. Eberhard's theorem) and on
the microscopic level (the hypothesis of local realism). Local realism
contradicts the mathematical structure of non-relativistic quantum
mechanics (NRQM). Because of that, the
violation of Bell's inequalities is often considered to be
experimental proof of the non-locality of quantum theory. However this is
not true, as there are two inseparable potential causes of violation
of the static Bell's and Wigner's inequalities: the absence of joint
Kolmogorov's probabilities for observables, and non-locality on
the microscopic level. To exclude the second possibility, it is necessary
to switch to calculations of probabilities and correlators in the
framework of QFT, which is local on the micro-level by definition, for
example because of Bogolyubov's principle of microcasuality
\cite{bogolyubov}.

However in QFT it is not possible to use static Bell's or Wigner's
inequalities, because it is not possible to exclude interactions of
quantum fields with each other and with vacuum fluctuations
\cite{Kazakov20122914}. I.e., in QFT any particle or system of
particles is an open system. Also the description of an entangled
quantum system at different spacetime points should take into
account relativistic effects, when the finite time of signal
propagation between the two parts of the entangled microsystem is
beyond the duration of macro-device response. This is undoubtedly true
for \cite{Nature526-spin-e-2015,PRL-2015-250401,
  PRL-2015-250402}. Hence to have the possibility of
a theoretical description of locality loophole-free
spatially separated experiments it is necessary to include time evolution
into  any Bell-like inequalities to assure their compatibility with
the theory of relativity.

The Wigner's inequalities are more suited for relativistic
generalization, because their intrinsic joint probabilities are well
defined both in NRQM and QFT. The correlators in the Bell's
inequalities are calculated from loop diagrams, part of which calculation depends
on the renormalization procedure at each order of perturbation theory. It
is not possible to get a definitive answer which does not depend on a
renormalization technique \cite{PeskinSchroeder}.

This paper is a logical continuation of a series 
\cite{Nikitin:2009sr, Nikitin:2014yaa,Nikitin:2015bca,Nikitin:2015mna}
in which we have studied possible relativistic corrections for static Wigner's
inequalities, and then introduced inequalities that generalized static
Wigner's inequalities for time-dependent ones required in
QFT. The main goal of the papers above was testing Bohr's
  complementarity principle in the relativistic domain. The complementarity
  principle is directly related to the concept of realism. We
  believe that the statement of ``testing the concept of realism'', most
  correctly reflects the gist and the results of
  \cite{Nikitin:2014yaa,Nikitin:2015bca,Nikitin:2015mna}. Also the
  connection between the violation of Leggett-Garg inequalities and the
  complementarity principle does not appear to be that obvious.
  In the current paper we will talk about realism, not
  complementarity, yet this approach is fully compatible with
Kolmogorov's axiomatics of probability theory and with the concept of
local realism; it is a particular realization of the supposition
of the independent time evolution of every physical (micro)system. It
would be nice to find a time evolution description that does not
require the supposition of independence. In
Section~\ref{section:mainBayes} of the current paper we will propose
such a description, which may be obtained in the framework of
Kolmogorov's probability theory using a bayesian approach and operates
only with (conditional) probabilities.

In addition to the works cited above, there are many proposals
to test Bell's inequalities in particle physics. Most of
these tests use oscillations of neutral pseudoscalar
mesons. They were discussed in the work \cite{Nikitin:2015bca}. 
Note also 
the analysis of the test of static Bell's inequalities in neutrino
oscillations \cite{sr-2013,}.

Time-dependent inequalities were proposed by A.~Leggett and A.~Garg in
their pioneering work \cite{Leggett:1985zz}. The initial goal of this
work was for testing whether quantum mechanics may be applied at
the macroscopic scale for many-particle quantum systems in a
coherent state. The inequality, which is
satisfied by two-times correlators of one observable, assumes that
this observable obeys the laws of classical physics (this is the concept
of ``macroscopic realism per se''. The expression of the
Leggett-Garg inequalities is similar to the Bell's inequalities in form
\cite{Clauser:1969ny}. Because of that the Leggett-Garg inequalities
are sometimes called ``temporal Bell inequalities''
\cite{PhysRevLett.107.090401}. This name is
not precise, because Bell's inequalities involve correlators of
various observables at one moment of time, while the Leggett-Garg
inequalities should involve different-time correlators of one
observable. The name ``time-dependent Bell's inequalities''
should be attributed to inequalities that contain
probabilities or correlators of various observables at different
moments of time. Following this logic, 
the ``time-dependent Wigner's inequalities''
were introduced \cite{Nikitin:2014yaa}. References to all the key works related to
Leggett-Garg inequalities may be found in the review \cite{repprogphys2014}.

In 2015 a successful experimental test of violation of the Leggett-Garg
inequalities were conducted: first in an experiment with non-classical
movement of a massive quantum particle over a lattice
\cite{ideal-2015}, and then in a violation of quantum
coherence in macroscopic crystals \cite{crystal}.

The Leggett-Garg inequalities may also be used to test for the existence of
joint probabilities of observables which have non-commuting operators
in NRQM, i.e. to test the hypothesis of realism. Let us note
that in the recent publication \cite{PhysRevLett.116.150401} authors
contested the well-known statement that the Leggett-Garg inequalities
might be applicable to test the principle of macroscopic
realism. But their applicability to test the principle of
(local) realism is not disputed. The Leggett-Garg
inequalities may be reproduced if one supposes that some hidden
parameters $\lambda$ whose probability density $\rho (\lambda)$ depends
on time in Markov's way, exist in a quantum
system \cite{PhysRevA.54.1798}. Such hidden parameters
automatically lead to non-negative joint probabilities. While a test of
macroscopic coherence requires soft measurements, a test of the existence of
the joint probabilities (i.e. the ``hypothesis of realism'' without
the term ``macroscopic'') requires the use of sets of parallel measurements, each
conducted at only two fixed moments of time
\cite{NatureCommunications2012}. This approach related to the 
methodology of the measurement of four distinct correlators in Bell's
inequalities, so one might expect that this approach would not be free
of the contextuality loophole. The use of projection measurements
and sets of parallel experiments for testing the hypothesis of realism
in particle physics is considered in
\cite{Gangopadhyay:2013aha}, where entangled pairs of pseudoscalar
mesons are used. In 2016 the neutrino experiment MINOS reported a test of
the Leggett-Garg inequalities in neutrino oscillations \cite{neutrino2016}.

The current paper considers some generic
experimental situations where one measures some properties of a
physical system; then constructs a Kolmogorov's space of elementary
outcomes and introduces into this space events, corresponding to these
experimental situations; we write some relations in terms of
conditional probabilities, because they are well-defined in both
classical and quantum physics (in contrast to joint probabilities);
and finally we show that 
corollaries of Bayes' theorem are violated for correlated quantum
systems.

There are many works dedicated to studies of
interconnections between the conditional probabilities in quantum and 
classical theories, starting from fundamental monography by
von~Neumann \cite{neumann1932} and paper by L\"uders
\cite{luders1951}, where rules for calculating conditional
probabilities were introduced. Important generalizations of the notion
of the conditional probability on a generalized probability space of
quantum mechanics were presented in 
\cite{Cassinelli1983,Cassinelli1984}. In the current paper the calculation
of conditional probabilities will be based on
\cite{Cassinelli1983,Cassinelli1984}. Generalizations of L\"uders' rule
for non-hermitian projection operators for entangled and open quantum
systems were proposed in
\cite{fuchs2001,fuchs2002}. Based on this generalization, a
quantum-bayesian interpretation of quantum mechanics was developed (so
called QBism) \cite{RevModPhys.85.1693,MerminNature2014}, which is
now, we believe, is one of the most elegant interpretations of quantum
theory. It provides a unified approach to classical and quantum
phenomena. QBism criticized in
\cite{Khrennikov2016}.

A versatile analysis of quantum conditional probability and its relation
to classics may be found in \cite{bobo-2013}. The main conclusion
of these works is that von~Neumann's formula can not be considered
a good generalization of classical conditional probability for
quantum phenomena, however there is no doubt in L\"uder's rule
\cite{luders1951} and its extention \cite{Cassinelli1983,Cassinelli1984}.

This paper consists of the following sections. In
Section~\ref{section:mainBayes}, 
using Kolmogorov's approach, conditional
probabilities, and Bayes' theorem, we obtain a static equality and a
time-dependent inequality, which allow us to test the hypothesis of realism
in time-dependent and open quantum systems. In
Section~\ref{section:spinInH} we use an example of correlated
spin-$1/2$ particles in an external stationary and homogeneous magnetic field to
demonstrate that in the framework of NRQM for open quantum systems the
relations obtained in Section~\ref{section:mainBayes} are
violated. Section~\ref{section:MbarM} is devoted to the study of
corollaries to violation of Bayes' theorem for systems of pseudoscalar
neutral mesons. Some experiments for testing the concept of
realism are proposed. They might be applicable to experiments
at the Large Hadron Collider~\cite{lhcb1,lhcb2,atlas,cms} and
Belle~II~\cite{belle2}. Appendices~\ref{sec:A}--\ref{sec:DD} contain
all the auxiliary formulae, that are required for derivation of the
results of Sections ~\ref{section:spinInH}--\ref{section:MbarM}.

\section{Testing the realism hypothesis using Bayes' theorem}
\label{section:mainBayes} 

Consider conditional probabilities
in classical and quantum theories using the observables $A$ and $B$.
In contrast to joint probabilities (\ref{w(ab)}), 
conditional probabilities like $w \left (b_{\beta'} \big | a_{\alpha'}
\right )$ are well-defined in both classical and quantum theories.

In the classical case the probability to measure value $b_{\beta'}$ of
the spectrum of observable $B$ assuming that value $a_{\alpha'}$ of
observable $A$ was measured can be written as follows:
\begin{eqnarray}
\label{ClassicalConditionalProb}
w \left (b_{\beta'} \big | a_{\alpha'} \right )\, =\, \frac{w(b_{\beta'} \cap a_{\alpha'})}{w(a_{\alpha'})}.
\end{eqnarray}
As was noted above, the joint probability $w(b_{\beta'} \cap
a_{\alpha'}) = w(a_{\alpha'} \cap b_{\beta'})$ always exists. Value
$w(a_{\alpha'}) \ne 0$ is the probability to measure value
$a_{\alpha'}$ from the spectrum of observable $A$. From
(\ref{ClassicalConditionalProb}) a simplest case of Bayes' theorem
can be derived:
\begin{eqnarray}
\label{BayesTheorem}
w \left (b_{\beta'} \big | a_{\alpha'} \right )\, w(a_{\alpha'})\, =\, w \left ( a_{\alpha'} \big | b_{\beta'}\right )\, w(b_{\beta'}),
\end{eqnarray}
where $w(b_{\beta'}) \ne 0$: this is the probability to measure value
$b_{\beta'}$ from the spectrum of observable $B$.

In the quantum case for the calculation of the conditional probability, one
should use von~Neumann's formula \cite{neumann1932}
\begin{eqnarray}
\label{neumann_rule}
 w(b_{\beta'}\,\vert\, a_{\alpha'})\,  =
\frac{\textrm{Tr} \left ( \hat P^{(B)}_{\beta'} \hat P^{(A)}_{\alpha'}\,\hat\rho_0\, \hat P^{(A)}_{\alpha'}  \hat P^{(B)}_{\beta'} \right )}{\textrm{Tr} \left ( \hat P^{(A)}_{\alpha'}\,\hat\rho_0\right )},
\end{eqnarray}
where $\hat\rho_0$ is the density matrix of the quantum system in the initial
state, $\hat P^{(A)}_{\alpha'}$ is the projector onto the state related to
the value $a_{\alpha'}$ of the spectrum of the observable $A$, and 
$\hat P^{(B)}_{\beta'}$ is the analogous projector for the value $b_{\beta'}$ of
the spectrum of the observable $B$. Applicability of
formula (\ref{neumann_rule}) does not require commutation of $\hat A$
and $\hat B$. It does not matter whether the state is
entangled or not (and hence to which subsystems the observables $A$ and
$B$ correspond). It also does not matter whether the quantum system is
open or not. It is obvious that if 
$
\left [ \hat A, \hat B\right ] \ne 0
$,
then
$$
\textrm{Tr} \left ( \hat P^{(B)}_{\beta'} \hat P^{(A)}_{\alpha'}\,\hat\rho_0\, \hat P^{(A)}_{\alpha'} \hat P^{(B)}_{\beta'} \right )\,\ne\,
\textrm{Tr} \left ( \hat P^{(A)}_{\alpha'} \hat P^{(B)}_{\beta'}\,\hat\rho_0\, \hat P^{(B)}_{\beta'} \hat P^{(A)}_{\alpha'} \right ).
$$
Hence in the quantum case it is not always possible to obtain an analog of
Bayes' theorem (\ref{BayesTheorem}). Moreover in the framework of QFT
and for open quantum systems, field operators and observable
operators, which consist of fields operators, do not commute at
distinct moments of time, i.e.  
$
\left [ \hat A(t_1),\, \hat  A(t_2) \right ] \ne 0
$.
Because of that the time can be treated as an additional degree of
freedom together with spatial directions $\vec a$, $\vec b$ and so on
(if we are talking about spins, for example).

Although we will refer to spin projections onto various
directions in space, the static equality and time-dependent
inequality obtained below are true for any set of dichotomic
observables.

Select three space directions, which are defined by
non-parallel unit vectors $\vec a$, $\vec b$, and $\vec c$. Let
the system of antifermion ``$1$'' and fermion ``$2$'' be in a singlet
spin state at time $t_0$. Suppose that the concept of
realism is true, i.e. spin projections $a^{(i)}_{\pm}$,
$b^{(i)}_{\pm}$, and $c^{(i)}_{\pm}$ of the antifermion and fermion onto
all three directions exist simultaneously at any time $t$, despite the
fact that they can not be measured by any macro-device. At the time
$t_0$ these projections obey the anticorrelation condition (\ref{pm=mp2}).

For this hypothetical situation it is easy to introduce a classical
probability model based on Kolmogorov's axiomatics. Let us define
space $\Omega$ of elementary outcomes $\omega_k$. Each of them is one
of the possible sets
$
\{
a^{(1)}_{\alpha}  b^{(1)}_{\beta}  c^{(1)}_{\gamma}\, 
a^{(2)}_{\alpha'} b^{(2)}_{\beta'} c^{(2)}_{\gamma'}
\}
$
of spin projections onto all three chosen directions $\vec a$, $\vec
b$, and $\vec c$, where indices 
$
\{\alpha, \alpha', \beta, \beta', \gamma, \gamma'\}  = \{ +, -\}
$. The set of the elements of the space $\Omega$ does not depend on time.

Denote ``elementary event''  $\mathcal{K}_{a^{(1)}_{\alpha}
  b^{(1)}_{\beta} c^{(1)}_{\gamma}\, a^{(2)}_{\alpha'}
  b^{(2)}_{\beta'} c^{(2)}_{\gamma'}}$ as a subset of all elementary
outcomes $\omega_k$ of the set $\Omega$
(i.e. 
$
\mathcal{K}_{a^{(1)}_{\alpha} b^{(1)}_{\beta} c^{(1)}_{\gamma}\, a^{(2)}_{\alpha'} b^{(2)}_{\beta'} c^{(2)}_{\gamma'}}\subseteq \Omega
$ 
and 
$
\omega_k \in \mathcal{K}_{a^{(1)}_{\alpha} b^{(1)}_{\beta} c^{(1)}_{\gamma}\, a^{(2)}_{\alpha'} b^{(2)}_{\beta'} c^{(2)}_{\gamma'}}
$
). Realization of any of these events gives rise to an element of
physical reality -- a concrete set of spin projections
$
\{ a^{(1)}_{\alpha} b^{(1)}_{\beta} c^{(1)}_{\gamma}\, a^{(2)}_{\alpha'} b^{(2)}_{\beta'} c^{(2)}_{\gamma'} \}
$.
The aggregate of the considered events forms an algebra
($\sigma$-algebra) $\mathcal{F}$. More complicated events may be
constructed by merging elementary events. 
It is possible to introduce a probability measure $w$ on $(\Omega,\,
\mathcal{F})$, which is always real and non-negative. It is
additive ($\sigma$-additive) for non-intersecting events. Using this
measure we can define probabilities of joint and conditional events on
$\Omega$.

In order to derive a static equality let us consider three events  
$\mathcal{S}_1 (t_0) = \{a^{(2)}_+, b^{(1)}_+\}$, $\mathcal{S}_2 (t_0)
= \{c^{(2)}_{+}, b^{(1)}_{+}\}$, and event $\mathcal{S}_3(t_0)$, when
the fermion-antifermion pair is in the spin singlet state at $t_0$. Events
$\mathcal{S}_1 (t_0)$ and $\mathcal{S}_2 (t_0)$ can be easily
constructed in $\Omega$ using the elementary events and condition
(\ref{pm=mp2}):
\begin{eqnarray}
\label{ABC}
\mathcal{S}_1(t_0) &=&
\mathcal{K}_{a^{(1)}_- b^{(1)}_+ c^{(1)}_+\,   a^{(2)}_+ b^{(2)}_- c^{(2)}_-}(t_0)\,\cup\,
\mathcal{K}_{a^{(1)}_- b^{(1)}_+ c^{(1)}_-\,  a^{(2)}_+ b^{(2)}_- c^{(2)}_+}(t_0),\nonumber \\ 
\mathcal{S}_2(t_0) &=& 
\mathcal{K}_{a^{(1)}_- b^{(1)}_+ c^{(1)}_-\,  a^{(2)}_+ b^{(2)}_- c^{(2)}_+}(t_0)\,\cup\,
\mathcal{K}_{a^{(1)}_+ b^{(1)}_+ c^{(1)}_-\, a^{(2)}_- b^{(2)}_- c^{(2)}_+}(t_0). \nonumber
\end{eqnarray}
In $\Omega$ space, event $\mathcal{S}_3$ is
defined as follows:
\begin{eqnarray}
\mathcal{S}_3 = \left\{ \left ( a^{(2)}_+, a^{(1)}_-\cup a^{(2)}_-, a^{(1)}_+\right ) \cup 
                     \left ( b^{(2)}_+, b^{(1)}_-\cup b^{(2)}_-, b^{(1)}_+\right ) \cup
                     \left ( c^{(2)}_+, c^{(1)}_-\cup c^{(2)}_-, c^{(1)}_+\right ) 
        \right\}.
\end{eqnarray}
This notation corresponds to the classical approach, which in this case is
identical to the concept of local realism, and in essence differs from
a description of event $\mathcal{S}_3 (t_0)$ in NRQM using a
maximally-entangled Bell state
\begin{eqnarray}
\label{psi-_t0}
\ket{\Psi^- (t_0)}\, =\,\frac{1}{\sqrt{2}}\,\left ( \ket{n^{(2)}_+}\ket{n^{(1)}_-}\, -\, \ket{n^{(2)}_-}\ket{n^{(1)}_+}\right ),
\end{eqnarray}
where $\vec n$ is any of the directions  $\vec a$, $\vec b$, or $\vec
c$. In QFT the initial state is defined using a Hamiltonian
(\ref{Heff_for_PS2ff}), which creates a corresponding entangled state
when calculating the evolution operator matrix element.

If the concept of realism is true, then we can consider non-negative
joint and conditional probabilities for the events $\mathcal{S}_1
(t_0)$, $\mathcal{S}_2 (t_0)$, and $\mathcal{S}_3 (t_0)$. It is possible
to apply to them a multiplication theorem on $(\Omega,\,
\mathcal{F})$. Then 
\begin{eqnarray}
w \left ( \mathcal{S}_1 \cap \mathcal{S}_2 |\mathcal{S}_3 \right ) &=& 
\frac{w \left (\mathcal{S}_1 \cap \mathcal{S}_2  \cap \mathcal{S}_3 \right )}{w \left (\mathcal{S}_3 \right )}\, =\,
\frac{w \left (\mathcal{S}_3 \cap \mathcal{S}_1 \cap \mathcal{S}_2  \right )}{w \left (\mathcal{S}_3 \right )}\, =\,
\frac{w \left (\mathcal{S}_3 \right )\, w \left ( \mathcal{S}_1 |\mathcal{S}_3 \right )\, w \left (\mathcal{S}_2 | \mathcal{S}_1 \cap \mathcal{S}_3  \right )}{w \left (\mathcal{S}_3 \right )}\,  \nonumber \\
&=& w \left ( \mathcal{S}_1 |\mathcal{S}_3 \right )\, w \left (\mathcal{S}_2 | \mathcal{S}_1 \cap \mathcal{S}_3  \right ). \nonumber
\end{eqnarray}
In analogy
\begin{eqnarray}
w \left ( \mathcal{S}_1 \cap \mathcal{S}_2 |\mathcal{S}_3 \right )\, =\, w \left ( \mathcal{S}_2 |\mathcal{S}_3 \right )\, w \left (\mathcal{S}_1 | \mathcal{S}_2 \cap \mathcal{S}_3  \right ). \nonumber
\end{eqnarray}
Equating these results with each other, we obtain the following variant
of Bayes' theorem: 
\begin{eqnarray}
\label{bayes-static-bell}
w \left ( \mathcal{S}_1\, |\, \mathcal{S}_3 \right )\, w \left ( \mathcal{S}_2\, |\,  \mathcal{S}_1 \cap \mathcal{S}_3 \right ) =  w \left ( \mathcal{S}_2\, |\, \mathcal{S}_3 \right )\, w \left ( \mathcal{S}_1\, |\,  \mathcal{S}_2 \cap \mathcal{S}_3 \right ).
\end{eqnarray}
Experiments that can test the hypothesis of realism using the static
equality (\ref{bayes-static-bell}) are fully identical to those
that test static Bell's
\cite{Bell:1964kc,Bell:1964fg,Clauser:1969ny} and Wigner's \cite{wigner}
inequalities. However it is easier to check the violation of
(\ref{bayes-static-bell})  than the violation of Bell's
inequalities. Also, (\ref{bayes-static-bell}) has an advantage over
the GHZ-equality \cite{GHZ1989,GHSZ1990, mermin1990}, because it
allows the experimenter to check in a system of only two (entangled) subsystems, while
an experimental check of violation of the GHZ-equality requires at least
three subsystems in an entangled state. The last condition makes it
almost impossible to study the GHZ-equality in particle physics.

The static equality (\ref{bayes-static-bell}) is one of two main
results of this paper. In Sections \ref{section:spinInH} and
\ref{section:MbarM}  it will be shown that this equality is violated
in the framework of quantum theory.

We now derive a time-dependent inequality which follows from the
hypothesis of realism and Bayes' theorem. In 
\cite{Nikitin:2014yaa,Nikitin:2015bca,Nikitin:2015mna} the authors
have suggested a variant of the time-dependent Wigner's inequality, based
on Kolmogorov's axiomatics and an assumption about the statistical
independence of processes of evolution of dichotomic variables of each
of two subsystems, which satisfy the condition (\ref{pm=mp2}) at the
time $t_0$ (and only at that time). For $t_1
\ne t_2$ this inequality may be written as follows

\resizebox{.82\linewidth}{!}{
\begin{minipage}{\linewidth}
{\begin{eqnarray}
\label{W-B-4t1t2}
&& w\left (a^{(2)}_+(t_2) \cap b^{(1)}_+ (t_1) \right )\,\le \\
&\le&
w\left (a^{(2)}_+(t_0) \to a^{(2)}_+ (t_2) \right )\,
\left (
w\left (b^{(1)}_+(t_0) \to b^{(1)}_+ (t_1) \right )\, +\, 
w\left (b^{(1)}_-(t_0) \to b^{(1)}_+ (t_1) \right )
\right )\, 
w\left (a^{(2)}_+(t_0)\cap c^{(1)}_+(t_0) \right )\, +\nonumber\\
&+&
w\left (a^{(2)}_-(t_0) \to a^{(2)}_+ (t_2) \right )\,
\left (
w\left (b^{(1)}_+(t_0) \to b^{(1)}_+ (t_1) \right )\, +\, 
w\left (b^{(1)}_-(t_0) \to b^{(1)}_+ (t_1) \right )
\right )\, 
w\left (a^{(2)}_-(t_0) \cap c^{(1)}_+(t_0) \right )\, +\nonumber\\
&+&
w\left (b^{(1)}_+(t_0) \to b^{(1)}_+ (t_1) \right )\,
\left (
w\left (a^{(2)}_+(t_0) \to a^{(2)}_+ (t_2) \right )\, +\, 
w\left (a^{(2)}_-(t_0) \to a^{(2)}_+ (t_2) \right )
\right )\,
w\left (c^{(2)}_+(t_0) \cap b^{(1)}_+(t_0) \right )\, + \nonumber\\ 
&+&
w\left (b^{(1)}_-(t_0) \to b^{(1)}_+ (t_1) \right )\,
\left (
w\left (a^{(2)}_+(t_0) \to a^{(2)}_+ (t_2) \right )\, +\, 
w\left (a^{(2)}_-(t_0) \to a^{(2)}_+ (t_2) \right )
\right )\,
w\left (c^{(2)}_+(t_0) \cap b^{(1)}_-(t_0) \right ),\nonumber 
\end{eqnarray}} 
\end{minipage}}

\noindent where the dichotomic variable of the first subsystem
is measured at the time $t_1 > t_0$, while the dichotomic variable of
the second subsystem is measured at the time $t_2 > t_0$. Although the
assumption of statistical independence of the evolution of classical
observables is almost obvious in the framework of the hypothesis of local
realism, it is quite hard to prove in some
cases. For this reason we would like to write an inequality in which the time
evolution is a consequence of a more common property of classical
objects rather than the property of statistical independence, which is
used in the derivation of  (\ref{W-B-4t1t2}). That more common
property might be Bayes' theorem.

Consider two moments in time: the initial $t_0$, and some $t >
t_0$. The anticorrelation condition
(\ref{pm=mp2}) is supposed to be true only at the time $t_0$. At any
other moment of time it might not be satisfied. As the space of 
elementary outcomes $\Omega$ does not depend on time, it is possible
to select the following events: the event $\mathcal{S}_1(t_0)  =
\{a^{(2)}_+, b^{(1)}_+, t_0\}$, the event $S_2(t) =
\{a^{(2)}_{\alpha'}, b^{(1)}_{\beta'}, t \}$, where $\{ \alpha',
\beta'\} = \{+,\, -\}$, and the event $S_3$ for which at the time $t_0$
the fermion-antifermion pair were in a singlet spin state. Under the
hypothesis of realism we again use the Bayes' theorem, but now for
the two moments of time.  
\begin{eqnarray}
w \Big ( \mathcal{S}_1 (t_0)\, \big |\, \mathcal{S}_3 (t_0) \Big )\, w \Big ( \mathcal{S}_2 (t)\, \big |\,  \mathcal{S}_1 (t_0) \cap \mathcal{S}_3 (t_0) \Big ) =  w \Big ( \mathcal{S}_2 (t)\, \big |\, \mathcal{S}_3 (t_0) \Big )\, w \Big ( \mathcal{S}_1 (t_0)\, \big |\,  \mathcal{S}_2 (t) \cap \mathcal{S}_3 (t_0) \Big ). \nonumber
\end{eqnarray}
In this formula the conditional probability $w \Big ( \mathcal{S}_1
(t_0)\, \big |\,  \mathcal{S}_2 (t) \cap \mathcal{S}_3 (t_0) \Big )$
is badly defined mathematically in both NRQM and QFT. However if we
suppose that the hypothesis of realism is true then this conditional
probability must satisty the following conditions:  
$$
0\,\le\, w \Big ( \mathcal{S}_1 (t_0)\, \big |\,  \mathcal{S}_2 (t) \cap \mathcal{S}_3 (t_0) \Big )\,\le\, 1.
$$ 
We obtain the time-dependent inequality: 
\begin{eqnarray}
\label{bayes-dinamic-bell}
w \Big ( \mathcal{S}_1 (t_0)\, \big |\, \mathcal{S}_3 (t_0) \Big )\, w \Big ( \mathcal{S}_2 (t)\, \big |\,  \mathcal{S}_1 (t_0) \cap \mathcal{S}_3 (t_0) \Big )\,
    \le\,   w \Big ( \mathcal{S}_2 (t) \big |\, \mathcal{S}_3 (t_0) \Big ). 
\end{eqnarray}
The time-dependent inequality (\ref{bayes-dinamic-bell}) is the second
main result of the paper. Only two directions, $\vec a$ and $\vec b$,
were used in the derivation of this inequality, not three or more as in
Bell's and Wigner's inequalities. Potentially that allows experimenter to test 
inequality (\ref{bayes-dinamic-bell}) using only one series of
experiments and thus evade the contextuality loophole.

It is suitable to write inequality
(\ref{bayes-dinamic-bell}) using the spin $1/2$ projections onto 
directions $\vec a$ and $\vec b$:
\begin{eqnarray}
\label{neravenstvobayesa}
&& w  \Big (\left \{ a^{(2)}_+, b^{(1)}_+, t_0 \right \}\,\big |\, \mathcal{S}_3 (t_0) \Big )\,\,\,
       w  \Big (\left \{a^{(2)}_{\alpha'}, b^{(1)}_{\beta'}, t \right \}\, \big | \, \left \{ a^{(2)}_+, b^{(1)}_+, t_0 \right \}\,\cap\, \mathcal{S}_3 (t_0) \Big ) \, \le\, \nonumber \\
&& \quad\quad\quad\quad\le\, w \Big (\left \{ a^{(2)}_{n'}, b^{(1)}_{m'}, t \right \}\,\big |\, \mathcal{S}_3 (t_0) \Big ),
\end{eqnarray}
where, let us emphasize again, we do not suppose any specific
dependence of the observables on time, and the event $\mathcal{S}_3
(t_0)$ may correspond, in principle, to any initial condition of a
system, not only the condition which satisfies 
(\ref{pm=mp2}).  Event $\mathcal{S}_1 (t_0)$ also
might be selected in a generic way as $\mathcal{S}_1(t_0)  =
\{a^{(2)}_{\alpha''}, b^{(1)}_{\beta''}, t_0\}$; however that
generalization does not lead to any new types of violation of the
inequality  (\ref{neravenstvobayesa}).

\section{Corollaries of Bayes' theorem and anticorrelated spins in an
  external magnetic field} 
\label{section:spinInH} 

We first show that the static equation (\ref{bayes-static-bell})
is violated in NRQM if we consider a positron and electron in a spin
singlet Bell state. The density matrix corresponding to this state is
$
\hat\rho_0 = \ket{\Psi^- (t_0)}\bra{\Psi^-(t_0)}
$. For tests of static inequality the time is not important, so let
$t_0 = 0$. Main formulae required for derivation of
conditional probabilities in static equation
(\ref{bayes-static-bell}) are given in Appendix~\ref{sec:A}. State
vector $\ket{\Psi^- (t_0)}$ is defined as  (\ref{correlation-t=0}).

In plane $(x,\, z)$, define three directions $\vec a$, $\vec
b$, and $\vec c$. Projectors onto the events $\mathcal{S}_1 (t_0) =
\{a^{(2)}_+, b^{(1)}_+\}$, $\mathcal{S}_2 (t_0) = \{c^{(2)}_{+},
b^{(1)}_{+}\}$, and spin singlet $\mathcal{S}_3(t_0)$ are 
\begin{eqnarray}
\label{P-t0}
\hat P_{\mathcal{S}_1}\, =\,\ket{a^{(2)}_+}\ket{b^{(1)}_+}\bra{b^{(1)}_+}\bra{a^{(2)}_+},\quad
\hat P_{\mathcal{S}_2}\, =\,\ket{c^{(2)}_+}\ket{b^{(1)}_+}\bra{b^{(1)}_+}\bra{c^{(2)}_+}\quad
\textrm{and}\quad
\hat P_{\mathcal{S}_3}\, =\,\hat \rho_0.
\end{eqnarray}
Using (\ref{P-t0}), (\ref{wf-t0}), and (\ref{neumann_rule}) we
obtain
\begin{eqnarray}
\label{w_S1-and-w_S2}
w \left ( \mathcal{S}_1\, |\, \mathcal{S}_3 \right ) & =& \left | \bracket{\Psi^- (t_0)\,}{a^{(2)}_+}\ket{b^{(1)}_+}\right |^2 =\,
\frac{1}{2}\,\sin^2\,\frac{\theta_{ab}}{2}, \\
w \left ( \mathcal{S}_2\, |\, \mathcal{S}_3 \right )  & =& \left | \bracket{\Psi^- (t_0)\,}{c^{(2)}_+}\ket{b^{(1)}_+}\right |^2 =\,
\frac{1}{2}\,\sin^2\,\frac{\theta_{bc}}{2}, \nonumber
\end{eqnarray}
where $\theta_{\alpha \beta} = \theta_\alpha - \theta_\beta$.

For calculation of the conditional probabilities $w \left (
  \mathcal{S}_2\, |\,  \mathcal{S}_1 \cap \mathcal{S}_3 \right )$ and
$w \left ( \mathcal{S}_1\, |\,  \mathcal{S}_2 \cap \mathcal{S}_3
\right )$ it is necessary to define projectors onto states 
$
\ket{\Psi_{ \mathcal{S}_1 \cap \mathcal{S}_3}}
$
and
$
\ket{\Psi_{ \mathcal{S}_2 \cap \mathcal{S}_3}}
$. In the general case such a procedure might be non-trivial. However
the isotropy of Bell state $\ket{\Psi^- (t_0)}$ allows us to obtain a
simple calculation algorithm. Let us find for example $\ket{\Psi_{
    \mathcal{S}_1 \cap \mathcal{S}_3}}$.  Rewrite
(\ref{psi-_t0}) in terms of the projection onto the direction $\vec a$:
\begin{eqnarray}
\ket{\Psi^- (t_0)} &=& \frac{1}{\sqrt{2}}\,\left ( \ket{a^{(2)}_+}\ket{a^{(1)}_-}\, -\, \ket{a^{(2)}_-}\ket{a^{(1)}_+}\right )\, = \nonumber \\
&=& \frac{1}{\sqrt{2}}\,\left ( \ket{a^{(2)}_+}\,\left [ - \sin\frac{\theta_{ab}}{2}\ket{b^{(1)}_+} + \cos\frac{\theta_{ab}}{2}\ket{b^{(1)}_-}  \right ]\, -\, \ket{a^{(2)}_-}\ket{a^{(1)}_+}\right )\, = \nonumber \\
&=& ...\, +\, \ket{\Psi_{ \mathcal{S}_1}}\, +\, ...\, .
\end{eqnarray}
From this, according to the superposition principle and the results of
\cite{Cassinelli1983,Cassinelli1984}, we obtain the
non-normalized state vector of event $ \mathcal{S}_1 \cap
\mathcal{S}_3$: 
$$
\ket{\Psi_{ \mathcal{S}_1 \cap \mathcal{S}_3}}\,
 =\, -\,\frac{1}{\sqrt{2}}\,\sin\,\frac{\theta_{ab}}{2}\,\ket{a^{(2)}_+}\ket{b^{(1)}_+}\, 
=\, -\,\frac{1}{\sqrt{2}}\,\sin\,\frac{\theta_{ab}}{2}\,\ket{\Psi_{ \mathcal{S}_1}}.
$$
Hence 
\begin{eqnarray}
\label{P_S1}
\hat P_{ \mathcal{S}_1 \cap \mathcal{S}_3}\, =\, \ket{\Psi_{ \mathcal{S}_1 \cap \mathcal{S}_3}}\bra{\Psi_{ \mathcal{S}_1 \cap \mathcal{S}_3}}\, =\,
\frac{1}{2}\,\sin^2\,\frac{\theta_{ab}}{2}\, \hat P_{\mathcal{S}_1}. \nonumber
\end{eqnarray}
In analogy
\begin{eqnarray}
\label{P_S2}
\hat P_{ \mathcal{S}_2 \cap \mathcal{S}_3}\, =\, 
\frac{1}{2}\,\sin^2\,\frac{\theta_{bc}}{2}\, \hat P_{\mathcal{S}_2}. \nonumber
\end{eqnarray}
Using the von~Neumann rule (\ref{neumann_rule}), we obtain the
following result. If the $e^+ e^-$-pair is in a spin singlet state, 
\begin{eqnarray}
\label{wS1S2}
w \left ( \mathcal{S}_2\, |\,  \mathcal{S}_1 \cap \mathcal{S}_3 \right ) &=&
\frac{\textrm{Tr} \left ( \hat P_{\mathcal{S}_2} \hat P_{ \mathcal{S}_1 \cap \mathcal{S}_3 }\,\hat\rho_0\, \hat P_{ \mathcal{S}_1 \cap \mathcal{S}_3 }  \hat P_{\mathcal{S}_2} \right )}{\textrm{Tr} \left ( \hat P_{ \mathcal{S}_1 \cap \mathcal{S}_3 }\,\hat\rho_0\,\hat P_{ \mathcal{S}_1 \cap \mathcal{S}_3 }\right )}\, =\,
\frac{\textrm{Tr} \left ( \hat P_{\mathcal{S}_2} \hat P_{ \mathcal{S}_1}\,\hat\rho_0\, \hat P_{ \mathcal{S}_1}  \hat P_{\mathcal{S}_2} \right )}{\textrm{Tr} \left ( \hat P_{ \mathcal{S}_1}\,\hat\rho_0\,\hat P_{ \mathcal{S}_1}\right )}\, = \\
&=&
\frac{\textrm{Tr} \left ( \hat P_{\mathcal{S}_2} \hat P_{ \mathcal{S}_1}  \hat P_{\mathcal{S}_2} \right )}{\textrm{Tr} \left ( \hat P_{ \mathcal{S}_1} \right )}\, =\,
\textrm{Tr} \left ( \hat P_{\mathcal{S}_1} \hat P_{\mathcal{S}_2} \right )\, =\,
w \left ( \mathcal{S}_1\, |\,  \mathcal{S}_2 \cap \mathcal{S}_3 \right ). \nonumber
\end{eqnarray}
We used the fact that for a Bell state $\hat\rho_0$,
\begin{eqnarray}
\label{proporc}
\hat P_{ \mathcal{S}_1}\,\hat\rho_0\, \hat P_{ \mathcal{S}_1}\,\sim\,\hat P_{ \mathcal{S}_1}\quad
\textrm{and}\quad
\hat P_{ \mathcal{S}_2}\,\hat\rho_0\, \hat P_{ \mathcal{S}_2}\,\sim\,\hat P_{ \mathcal{S}_2}. 
\end{eqnarray}
The equality of the conditional probabilities
(\ref{wS1S2}) is not general and is only related to the special choice of
the initial state $\ket{\Psi^-(t_0)}$.

Substituting (\ref{w_S1-and-w_S2}) and (\ref{wS1S2}) into
(\ref{bayes-static-bell}), if NRQM is compatible with the
hypothesis of local realism (and with probability theory in
Kolmogorov's axiomatics), then for any three directions $\vec a$, $\vec
b$, and $\vec c$ in plane $(x,\, z)$, the following equation
should be always satisfied:
\begin{eqnarray}
\label{ab=bc}
\sin^2\,\frac{\theta_{ab}}{2}\, =\, \sin^2\,\frac{\theta_{bc}}{2}.
\end{eqnarray}
Obviously this is not true. If the vector $\vec a$ is
perpendicular to the vector $\vec b$, while the vector $\vec c$ is the 
bisector of the angle between $\vec a$ and $\vec b$, then (\ref{ab=bc})
is violated.

We now show that the time-dependent inequality
(\ref{bayes-dinamic-bell}) may also be violated in NRQM. 
Again consider an $e^+e^-$-pair, which at the time $t_o = 0$ is described
by the density matrix
$
\hat\rho_0 = \ket{\Psi^- (t_0)}\bra{\Psi^-(t_0)}
$. 
Put the system into an external constant and homogeneous
magnetic field $\vec{\mathcal{H}}$ aligned along the $y$-axis. From
all possible decays, select only those where the leptons are
propagated along the magnetic field. This is assumed for
simplification of calculation of probabilities. Choose two
space directions $\vec a$ and $\vec b$ lying in the plane $(x,
z)$; then it is more suitable to test the violation
of (\ref{neravenstvobayesa}) than of (\ref{bayes-dinamic-bell}).
Projectors onto the events $\mathcal{S}_1 (t_0) = \{a^{(2)}_+,
b^{(1)}_+\}$, and $\mathcal{S}_3(t_0)$ may be written as
\begin{eqnarray}
\label{P-t00}
\hat P_{\mathcal{S}_1}\, =\,\ket{a^{(2)}_+}\ket{b^{(1)}_+}\bra{b^{(1)}_+}\bra{a^{(2)}_+},\quad
\textrm{and}\quad
\hat P_{\mathcal{S}_3}\, =\,\hat \rho_0,
\end{eqnarray}
The expression for the conditional probability 
$
w \Big ( \mathcal{S}_1 (t_0)\, \big |\, \mathcal{S}_3 (t_0) \Big )\, =\, w  \Big (\left \{ a^{(2)}_+, b^{(1)}_+, t_0 \right \}\,\big |\, \mathcal{S}_3 (t_0) \Big )
$ is calculated in (\ref{w_S1-and-w_S2}). In order to obtain two other
conditional probabilities, which are included in the formulae
(\ref{bayes-dinamic-bell}) and (\ref{neravenstvobayesa}), it is
necessary to consider four cases for different values of  $\alpha'$
and $\beta'$ for the event $\mathcal{S}_2(t)$.

\textbf{a)} Let at time $t > t_0$ the indices $\alpha' = +$ and $\beta'
= +$, i.e. $\mathcal{S}_2 (t) = \{a^{(2)}_+, b^{(1)}_+\}$. Considering that
$\hat \rho^2_0 = \hat\rho_0$ and that $\textrm{Tr}\,\hat\rho_0 =1$,
using the von~Neumann rule (\ref{neumann_rule}), we find that 
\begin{eqnarray}
w \Big ( \mathcal{S}_2 (t)\, \big |\, \mathcal{S}_3 (t_0) \Big ) & =& w  \Big (\left \{ a^{(2)}_+, b^{(1)}_+, t\right \}\,\big |\, \mathcal{S}_3 (t_0) \Big )\,=\,
\frac{\textrm{Tr} \left ( \hat P_{\mathcal{S}_2} \hat U(t, t_0) \hat P_{ \mathcal{S}_3}\,\hat\rho_0\, \hat P_{ \mathcal{S}_3} \hat U^{\dagger}(t, t_0)  \hat P_{\mathcal{S}_2} \right )}{\textrm{Tr} \left ( \hat P_{ \mathcal{S}_3}\,\hat\rho_0\,\hat P_{ \mathcal{S}_3}\right )}\, = \nonumber \\
&=& \textrm{Tr} \left ( \hat P_{\mathcal{S}_2} \hat U(t, t_0)\,\hat\rho_0\,\hat U^{\dagger}(t, t_0)  \hat P_{\mathcal{S}_2} \right )\, =\,
\left | \bracket{\Psi^- (t)\,}{a^{(2)}_+}\ket{b^{(1)}_+}\right |^2. \nonumber
\end{eqnarray}
The square of the modulus of the corresponding matrix element is calculated
using expressions (\ref{wf-t-e-}), (\ref{wf-t-e+}), and (\ref{psi-_t}). Finally:
\begin{eqnarray}
\label{S2(t)++}
w \Big ( \mathcal{S}_2 (t)\, \big |\, \mathcal{S}_3 (t_0) \Big ) \, =\, w  \Big (\left \{ a^{(2)}_+, b^{(1)}_+, t\right \}\,\big |\, \mathcal{S}_3 (t_0) \Big )\,=\,
\frac{1}{2}\,\sin^2
  \left (
             \frac{\theta_{ba}}{2}\, +\, 2 \omega t
  \right ).
\end{eqnarray}
Using the von~Neumann rule, property (\ref{proporc}), and formulae
(\ref{wf-t-e-}) and (\ref{wf-t-e+}), one can obtain
\begin{eqnarray}
\label{S2(t)++S1S3}
&& w \Big ( \mathcal{S}_2 (t)\, \big |\,  \mathcal{S}_1(t_0) \cap \mathcal{S}_3(t_0) \Big ) = 
w  \Big (\left \{a^{(2)}_{+}, b^{(1)}_{+}, t \right \}\, \big | \, \left \{ a^{(2)}_+, b^{(1)}_+, t_0 \right \}\,\cap\, \mathcal{S}_3 (t_0) \Big )\, = \\
&=& \frac{\textrm{Tr} \left ( \hat P_{\mathcal{S}_2} \hat U(t, t_0) \hat P_{\mathcal{S}_1 \cap \mathcal{S}_3}\,\hat\rho_0\, \hat P_{\mathcal{S}_1 \cap\mathcal{S}_3} \hat U^{\dagger}(t, t_0)  \hat P_{\mathcal{S}_2} \right )}{\textrm{Tr} \left ( \hat P_{\mathcal{S}_1 \cap\mathcal{S}_3}\,\hat\rho_0\,\hat P_{\mathcal{S}_1 \cap\mathcal{S}_3}\right )}\, =\,
  \frac{\textrm{Tr} \left ( \hat P_{\mathcal{S}_2} \hat U(t, t_0) \hat P_{\mathcal{S}_1}\,\hat\rho_0\, \hat P_{\mathcal{S}_1} \hat U^{\dagger}(t, t_0)  \hat P_{\mathcal{S}_2} \right )}{\textrm{Tr} \left ( \hat P_{\mathcal{S}_1}\,\hat\rho_0\,\hat P_{\mathcal{S}_1}\right )}\, = \nonumber \\
&=& \frac{\textrm{Tr} \left ( \hat P_{\mathcal{S}_2} \hat U(t, t_0) \hat P_{\mathcal{S}_1}\,\hat U^{\dagger}(t, t_0)  \hat P_{\mathcal{S}_2} \right )}{\textrm{Tr} \left ( \hat P_{\mathcal{S}_1}\right )}\, =\,
\Big | \bracket{a^{(2)}_{+}}{a^{(2)}_{+} (t)} \Big |^2 \Big | \bracket{b^{(1)}_{+}}{b^{(1)}_{+} (t)} \Big |^2 =\, 
\cos^4 \left ( \omega\, t\right ). \nonumber
\end{eqnarray}
We have used the standard
properties of projection operators: $\hat P_{\mathcal{S}_i}^2 = \hat
P_{\mathcal{S}_i}$ and  $\textrm{Tr}\,\hat P_{\mathcal{S}_i} =1$,
where $i = \{1, 2\}$.

Combine results (\ref{w_S1-and-w_S2}),
(\ref{S2(t)++}), and (\ref{S2(t)++S1S3}), and substitute them into the
inequality (\ref{neravenstvobayesa}), to get  
\begin{eqnarray}
\label{neravenstvobayesa++}
\sin^2 \left ( \frac{\theta_{ba}}{2}\right )\,\cos^4 \left ( \omega\, t\right )\,\le\,\sin^2 \left ( \frac{\theta_{ba}}{2}\, +\, 2 \omega t \right ).
\end{eqnarray}
If the concept of realism is true then the inequality
(\ref{neravenstvobayesa++}) should never be violated. However if we
choose $t$ such that $\omega t = -
\theta_{ba}/4$, then (\ref{neravenstvobayesa++}) becomes
\begin{eqnarray}
\label{violate1}
\sin^2 \left ( \frac{\theta_{ba}}{2}\right )\,\cos^4 \left ( \frac{\theta_{ba}}{4}\right )\,\le\, 0,
\end{eqnarray}
which is violated for most angles $\theta_{ba}$. 

The inequality (\ref{neravenstvobayesa++}) may be tested 
selecting events from one experiment without changing
the internal state of a macro-device, to avoid the
contextuality loophole. However the detection loophole in this
case might still be open.

\textbf{b)} 
Let at the time $t > t_0$ the indices $\alpha'
= -$ and $\beta' = -$, i.e. let us consider the event $\mathcal{S}_2 (t)
= \{a^{(2)}_-, b^{(1)}_-\}$. Performing calculations analogous to
the above, we find the following values for the conditional probabilities:
\begin{eqnarray}
\label{S2(t)--S1S3}
w \Big ( \mathcal{S}_2 (t)\, \big |\, \mathcal{S}_3 (t_0) \Big ) &=& w  \Big (\left \{ a^{(2)}_-, b^{(1)}_-, t\right \}\,\big |\, \mathcal{S}_3 (t_0) \Big )\,=\,
\frac{1}{2}\,\sin^2
  \left (
             \frac{\theta_{ba}}{2}\, +\, 2 \omega t
  \right ); \\
w \Big ( \mathcal{S}_2 (t)\, \big |\,  \mathcal{S}_1(t_0) \cap \mathcal{S}_3(t_0) \Big ) &=& 
w  \Big (\left \{a^{(2)}_{-}, b^{(1)}_{-}, t \right \}\, \big | \, \left \{ a^{(2)}_+, b^{(1)}_+, t_0 \right \}\,\cap\, \mathcal{S}_3 (t_0) \Big )\, = 
\sin^4 \left ( \omega\, t\right ). \nonumber
\end{eqnarray}
Substituting probabilities from (\ref{w_S1-and-w_S2}) and
(\ref{S2(t)--S1S3}) into inequality (\ref{neravenstvobayesa}), we
find that  
\begin{eqnarray}
\label{neravenstvobayesa--}
\sin^2 \left ( \frac{\theta_{ba}}{2}\right )\,\sin^4 \left ( \omega\, t\right )\,\le\,\sin^2 \left ( \frac{\theta_{ba}}{2}\, +\, 2 \omega t \right ).
\end{eqnarray}
If we choose $\omega t = - \theta_{ba}/4$, inequality
(\ref{neravenstvobayesa--}) is almost always violated:
\begin{eqnarray}
\label{violate2}
\sin^2 \left ( \frac{\theta_{ba}}{2}\right )\,\sin^4 \left ( \frac{\theta_{ba}}{4}\right )\,\le\, 0.
\end{eqnarray}

\textbf{c)} Finally let us consider the situation when  $\alpha' =
\mp$ and $\beta' = \pm$. Then $\mathcal{S}_2 (t) = \{a^{(2)}_\mp,
b^{(1)}_\pm\}$. The corresponding conditional probabilities are equal
to 
\begin{eqnarray}
\label{S2(t)+-S1S3}
w \Big ( \mathcal{S}_2 (t)\, \big |\, \mathcal{S}_3 (t_0) \Big ) &=& w  \Big (\left \{ a^{(2)}_\mp, b^{(1)}_\pm, t\right \}\,\big |\, \mathcal{S}_3 (t_0) \Big )\,=\,
\frac{1}{2}\,\cos^2
  \left (
             \frac{\theta_{ba}}{2}\, +\, 2 \omega t
  \right ); \\
w \Big ( \mathcal{S}_2 (t)\, \big |\,  \mathcal{S}_1(t_0) \cap \mathcal{S}_3(t_0) \Big ) &=& 
w  \Big (\left \{a^{(2)}_{\mp}, b^{(1)}_{\pm}, t \right \}\, \big | \, \left \{ a^{(2)}_+, b^{(1)}_+, t_0 \right \}\,\cap\, \mathcal{S}_3 (t_0) \Big )\, = 
\sin^2 \left ( \omega\, t\right )\,\cos^2 \left ( \omega\, t\right ). \nonumber
\end{eqnarray}
Inequality (\ref{neravenstvobayesa}) turns into 
\begin{eqnarray}
\label{neravenstvobayesa+-}
\sin^2 \left ( \frac{\theta_{ba}}{2}\right )\,\sin^2 \left ( \omega\, t\right )\,\cos^2 \left ( \omega\, t\right )\,\le\,\cos^2 \left ( \frac{\theta_{ba}}{2}\, +\, 2 \omega t \right ).
\end{eqnarray}
If we set $\omega t = \pi/4 - \theta_{ba}/4$, we obtain:
\begin{eqnarray}
\label{violate3}
\sin^2 \theta_{ba}\,\le\, 0,
\end{eqnarray}
which like (\ref{violate1}) and
(\ref{violate2}), is true for almost all 
choices of directions $\vec a$ and $\vec b$.

\section{Corollaries of Bayes' theorem for systems of neutral
  pseudoscalar mesons}
\label{section:MbarM} 

In Section~\ref{section:mainBayes}, equality
(\ref{bayes-static-bell}) and inequality
(\ref{bayes-dinamic-bell}) were obtained in terms of spin $1/2$
projections onto various directions in three-dimensional
space. Actually, relations (\ref{bayes-static-bell}) and
(\ref{bayes-dinamic-bell}) are true for any dichotomic observables of
any nature in any space.
In the case of neutral pseudoscalar mesons $M = \{ K,\, D,\, B_d,\, B_s
\}$, the dichotomic variables can be the flavour of the meson, its
$CP$-parity, and the lifetime (or mass). Pseudoscalar mesons are
unstable particles, hence they may
serve as a simple model of an open quantum system.
Formulae (\ref{bayes-static-bell}) and (\ref{bayes-dinamic-bell}), in
principle, may be tested in experiments at the Large Hadron Collider
\cite{lhcb1,lhcb2,atlas,cms}, at $B$-factory Belle~II \cite{belle2} and at
$\phi$-factories.

When decaying a neutral vector meson with the quantum numbers 
$J^{P\, C} = 1^{-\, -}$ of a photon into a $M \bar M$-pair, the latter rests in
the Bell state $\ket{\Psi^-}$ by flavour, $CP$-parity, or lifetime
($H/L$). In Appendix \ref{sec:B} the main properties of pseudoscalar
mesons are presented as well as the formula for evolution of an entangled
Bell state (here and below we use $\hbar = c =1$).

As it is impossible to unambiguously relate the spin projections to
the directions $\vec a$, $\vec b$, and $\vec c$ and projections of 
states of pseudoscalar mesons onto ``directions'' of flavour,
$CP$-parity and ``directions'' with a definitive mass/lifetime, we
need to consider some variants of these
correspondences. Note that 
$
\bracket{M}{\bar M} = \bracket{M_1}{M_2} = 0
$,
but
$
\bracket{M_H}{M_L} \ne 0
$.
Then it is suitable to use the following: $a_+ \to M$, $a_- \to \bar
M$, $b_+ \to M_L$, $b_- \to M_H$, $c_+ \to M_2$, and $c_- \to M_1$,
for the algorithms of calculation of projectors $\hat P_{
  \mathcal{S}_1 \cap \mathcal{S}_3}$ and $\hat P_{ \mathcal{S}_2 \cap
  \mathcal{S}_3}$ in Section~\ref{section:spinInH} to be analogous to
the ones from Section~\ref{section:MbarM}.

We now show that equality (\ref{bayes-static-bell}) is
violated in systems of neutral pseudoscalar mesons. At $t_0 = 0$ the
state of the $M\bar M$-system is defined by density matrix $\hat\rho_0 =
\ket{\Psi^- (t_0)}\bra{\Psi^-(t_0)}$, where the Bell state
$\ket{\Psi^- (t_0)}$ is defined by formula
(\ref{correlationBbarB-t=0}). Projection operators onto events
$\mathcal{S}_1 (t_0) = \{M^{(2)}, M^{(1)}_L\}$, $\mathcal{S}_2 (t_0) =
\{M^{(2)}_2, M^{(1)}_L\}$, and singlet state in the flavour space
$\mathcal{S}_3(t_0)$ are 
\begin{eqnarray}
\label{PM-t0}
\hat P_{\mathcal{S}_1} = \ket{M^{(2)}}\ket{M^{(1)}_L}\bra{M^{(1)}_L}\bra{M^{(2)}},\,
\hat P_{\mathcal{S}_2} = \ket{M^{(2)}_2}\ket{M^{(1)}_L}\bra{M^{(1)}_L}\bra{M^{(2)}_2}\,\,
\textrm{anf}\,\,
\hat P_{\mathcal{S}_3}\, =\,\hat \rho_0.
\end{eqnarray}
Using von~Neumann rule (\ref{neumann_rule}), formulae (\ref{PM-t0}),
and (\ref{correlationBbarB-t=0}) analogous to (\ref{w_S1-and-w_S2}),
one may write that (see formula \ref{w-BbarB-I})
\begin{eqnarray}
\label{w_S1M-and-w_S2M}
w \left ( \mathcal{S}_1\, |\, \mathcal{S}_3 \right ) & =& w \Big ( M^{(2)},\, M_L^{(1)},\,  t_0 \Big ) =\,
\frac{1}{2}\,\left | q \right |^2, \\
w \left ( \mathcal{S}_2\, |\, \mathcal{S}_3 \right )  & =& w \Big ( M_2^{(2)},\, M_L^{(1)},\,  t_0 \Big ) =\,
\frac{1}{4}\,\left | p + q \right |^2  \nonumber
\end{eqnarray}
Using a condition of orthogonality analogous to calculations from
Section \ref{section:spinInH} for non-normalized projection operators
onto the states corresponding to events $\hat P_{ \mathcal{S}_1 \cap
  \mathcal{S}_3}$ and $\hat P_{ \mathcal{S}_2 \cap \mathcal{S}_3}$, we have
\begin{eqnarray}
\label{P_S1MS2M}
\hat P_{ \mathcal{S}_1 \cap \mathcal{S}_3}\, =\, \frac{1}{8 \left | q \right |^2}\, \hat P_{\mathcal{S}_1}, \quad
\hat P_{ \mathcal{S}_2 \cap \mathcal{S}_3}\, =\, \frac{\left | p + q \right |^2}{16\,\left | p\, q \right |^2}\, \hat P_{\mathcal{S}_2}. \nonumber
\end{eqnarray}
Then calculation of conditional probabilities according to
von~Neumann's rule (\ref{neumann_rule}) leads to the equality
\begin{eqnarray}
\label{wS1S2M}
w \left ( \mathcal{S}_2\, |\,  \mathcal{S}_1 \cap \mathcal{S}_3 \right )\, =\,
w \left ( \mathcal{S}_1\, |\,  \mathcal{S}_2 \cap \mathcal{S}_3 \right )\, =\,
\textrm{Tr} \left ( \hat P_{\mathcal{S}_1} \hat P_{\mathcal{S}_2} \right )\, =\,\frac{1}{2},
\end{eqnarray}
which is analogous to formula (\ref{wS1S2}). Substituting
(\ref{w_S1M-and-w_S2M}) and (\ref{wS1S2M}) into the equality
(\ref{bayes-static-bell}), we have 
\begin{eqnarray}
\label{2=0M01}
2\, =\, \left | 1\, +\, \frac{p}{q} \right |^2.
\end{eqnarray}
The equality should be satisfied if the hypothesis of realism is
true. As shown in Appendix \ref{sec:B} for neutral $K$-- and
$D$--mesons, the ratio $q/p$ is close to $+1$ while for $B_d$-- and
$B_s$--mesons it is almost always equals to $-1$. Hence (\ref{2=0M01})
implies false relations like $2 \approx 4$ and $2 \approx 0$.

If we use the correspondence  $a_+ \to M$, $a_- \to \bar M$, $b_+ \to
M_H$, $b_- \to M_L$, $c_+ \to M_2$ and $c_- \to M_1$, which differs by
interchanging $M_L$ and $M_H$, then in the framework of the
hypothesis of realism we come to the following equality:
\begin{eqnarray}
\label{2=0M02}
2\, =\, \left | 1\, -\, \frac{p}{q} \right |^2,
\end{eqnarray}
which, like (\ref{2=0M01}), is not true in
flavour-entangled systems of neutral pseudoscalar mesons. Examination
of other correspondences leads to equalities which do not provide
anything new.  %

Now consider the time-dependent inequality
(\ref{bayes-dinamic-bell}), and let us show that it is also violated in
systems of neutral pseudoscalar mesons. The
most natural choice of ``directions'' for neutral $D$-- and
$B_q$--mesons is related to states with a definite flavour and
$CP$--parity. For example at $B$--factories the flavour of neutral
$B_d$--meson is determined by a lepton sign in semileptonic decay,
while $CP$--parity is determined using the decay $B_d \to J/\psi K^0_S$. At
hadron machines the task is much more complicated, as $B\bar
B$--pairs are mainly produced not through the $\Upsilon (4S)$ decay
but through the process of direct hadronization of $b\bar b$ quark
pairs. In order to select states corresponding to $\Upsilon (4S)$, one
needs to know the invariant mass of the $B \bar B$--pair, i.e. to fully
reconstruct the energy and momentum of each $B_d$--meson. So at hadron
machines it is not possible to use semileptonic decays with branching ratios
of the order of $10^{-1}$ for determination of the $B$-meson
flavour. An alternative way to detect the flavour is 
to use the cascade decay $B^0_d \to (D^- \to K^- K^+ \pi^-)K^+$,
with a branching ratio of about $10^{-5}$. For $B_s$--mesons at the LHC
experiments, the flavour can be determined in the decay $B^0_s \to (D^-_s
\to K^- \pi^+ \pi^-) \pi^+$. At hadron machines, the statistics
required for testing for violation of (\ref{bayes-dinamic-bell})
should be higher by a few orders of magnitude than at the
$B$--factories. For $D$--mesons the situation is slightly better because
the flavour of the $D$--meson may be determined in the decay $D^0 \to K^-
\pi^+$, which has a branching ratio of about 4\%. In the case of $K$--mesons, the
hadron machines are not suitable at all, and the test for
violation of the inequality (\ref{bayes-dinamic-bell}) must be
performed only at $\phi$-factories.

Consider events $\mathcal{S}_1 (t_0) = \{M_1^{(2)}, M^{(1)}\}$,
$\mathcal{S}_2 (t) = \{M^{(2)}_1, M^{(1)}\}$, and event
$\mathcal{S}_3(t_0)$, which corresponds to the singlet spin state of
the $M \bar M$-pair. Using formulae (\ref{w-BbarB-I}),
(\ref{w-BbarB-II}) from Appendix \ref{sec:D} and the algorithm for
calculation of projector $\hat P_{ \mathcal{S}_1 \cap \mathcal{S}_3}$
described in Section \ref{section:spinInH}, we find 
\begin{eqnarray}
\label{probM1M}
w \Big ( \mathcal{S}_1 (t_0)\, \big |\, \mathcal{S}_3 (t_0) \Big ) &=& w(M_1^{(2)},\, M^{(1)},\, t_0)\,=\, \frac{1}{4}; \nonumber \\
w \Big ( \mathcal{S}_2 (t)\, \big |\, \mathcal{S}_3 (t_0) \Big ) &=& w(M_1^{(2)},\, M^{(1)},\, t)\,=\, \frac{1}{4}\, e^{-\, 2 \Gamma\, t}\\
w \Big ( \mathcal{S}_2 (t)\,  \big |\,  \mathcal{S}_1(t_0) \cap \mathcal{S}_3(t_0) \Big ) &=& w(M_1(0) \to M_1(t))\, w(M (0) \to M (t))\, =  \nonumber\\
    &=&
     \left | 
              g_+(t)\, -\, \frac{1}{2}\,\left ( \frac{q}{p} + \frac{p}{q}\right )\, g_-(t)
     \right |^2\, |g_+ (t)|^2.
\nonumber
\end{eqnarray}
Using the set of probabilities (\ref{probM1M}), inequality
(\ref{bayes-dinamic-bell}) becomes
\begin{eqnarray}
\label{timeM1M}
     \left | 
              g_+(t)\, -\, \frac{1}{2}\,\left ( \frac{q}{p} + \frac{p}{q}\right )\, g_-(t)
     \right |^2\, |g_+ (t)|^2\,\, e^{2 \Gamma\, t}\,\le\, 1.
\end{eqnarray}
To understand for which mesons inequality (\ref{timeM1M})
may be violated, consider the case without
$CP$--violation, i.e when $q/p = \pm 1$. Then for $K$-- and $D$--mesons
formula (\ref{timeM1M}) becomes
$$
e^{\Delta \Gamma\, t/2}\, +\, 2\,\cos \Big ( \Delta M\, t\Big )\,\le\, 3\, e^{- \Delta \Gamma\, t/2},
$$
which is not violated for any values of $t \ge 0$, as $\Delta \Gamma <
0$ (see Table \ref{table:parameters}). For $B_q$-mesons, inequality
(\ref{timeM1M}) becomes
\begin{eqnarray}
\label{timeM1M-simple}
e^{-\Delta \Gamma\, t/2}\, +\, 2\,\cos \Big ( \Delta M\, t\Big )\,\le\, 3\, e^{\Delta \Gamma\, t/2}.
\end{eqnarray}
We estimate, for $B^0_s$-mesons, at which times $t \ge 0$ the
inequality (\ref{timeM1M-simple}) should be violated. From Table
\ref{table:parameters} one can see that 
$
\cos \Big ( \Delta M\, t\Big ) \approx \cos \Big ( 200\, \Delta \Gamma\, t\Big )
$. Hence for small adjustments of parameter $t$, the argument of the cosine
goes through its full period. Hence the condition of guaranteed violation of inequality
(\ref{timeM1M-simple}) is given by 
$$
e^{-\Delta \Gamma\, t/2}\,\ge\, 3\, e^{\Delta \Gamma\, t/2}\, +\, 2\quad \textrm{or}\quad t\,\ge\, \frac{2 \ln 3}{|\Delta \Gamma|}.
$$

$|\Delta \Gamma_{B_s}| = (0.091 \pm 0.008) \times 10^{12}$~s$^{-1}$
and $\tau_{B_s} = (1.512 \pm 0.007) \times 10^{-12}$~s
\cite{Beringer:1900zz}, where $\tau_{B_s}$ is the average lifetime of
the $B^0_s$--meson, so the condition for guaranteed violation of
inequality (\ref{timeM1M-simple}) is $t \ge 16 \tau_{B_s}$. Due to the
small magnitude of $CP$--violation effects, the exact inequality
(\ref{timeM1M}) should be violated at times of the same order.

From the calculations above it is clear that for various choices of
events $\mathcal{S}_1 (t_0)$ and $\mathcal{S}_2 (t)$, inequality
(\ref{bayes-dinamic-bell}) will always transform into the following
expression: 
\begin{eqnarray}
\label{F_N}
\textrm{F}_{N } (x,\, r,\,\zeta,\,\lambda)\,\le\, 1,
\end{eqnarray}
where functions $F_N$ depend on dimentionless variables $x = \Delta
\Gamma\, t$, $\lambda = \Delta M/ \Delta \Gamma$, the modulus of  $r$ and
phase $\zeta$ of the ratio $q/p$ (see Appendix \ref{sec:B}). 

There are some experimental limits on the range of possible values of
parameters $r$ and $\zeta$. In \cite{Nikitin:2015bca} and
\cite{Nikitin:2015mna} concerning the modelling of the violation of
time-dependent Wigner's inequalities in systems of $B_s$-mesons
the following values were selected: $r = 1.004$ and $\zeta =
185^{\textrm{o}}$. In the present paper we will also use these values.

\begin{figure}[bt]
	\begin{tabular}{cc}
         \includegraphics[width=7.9cm]{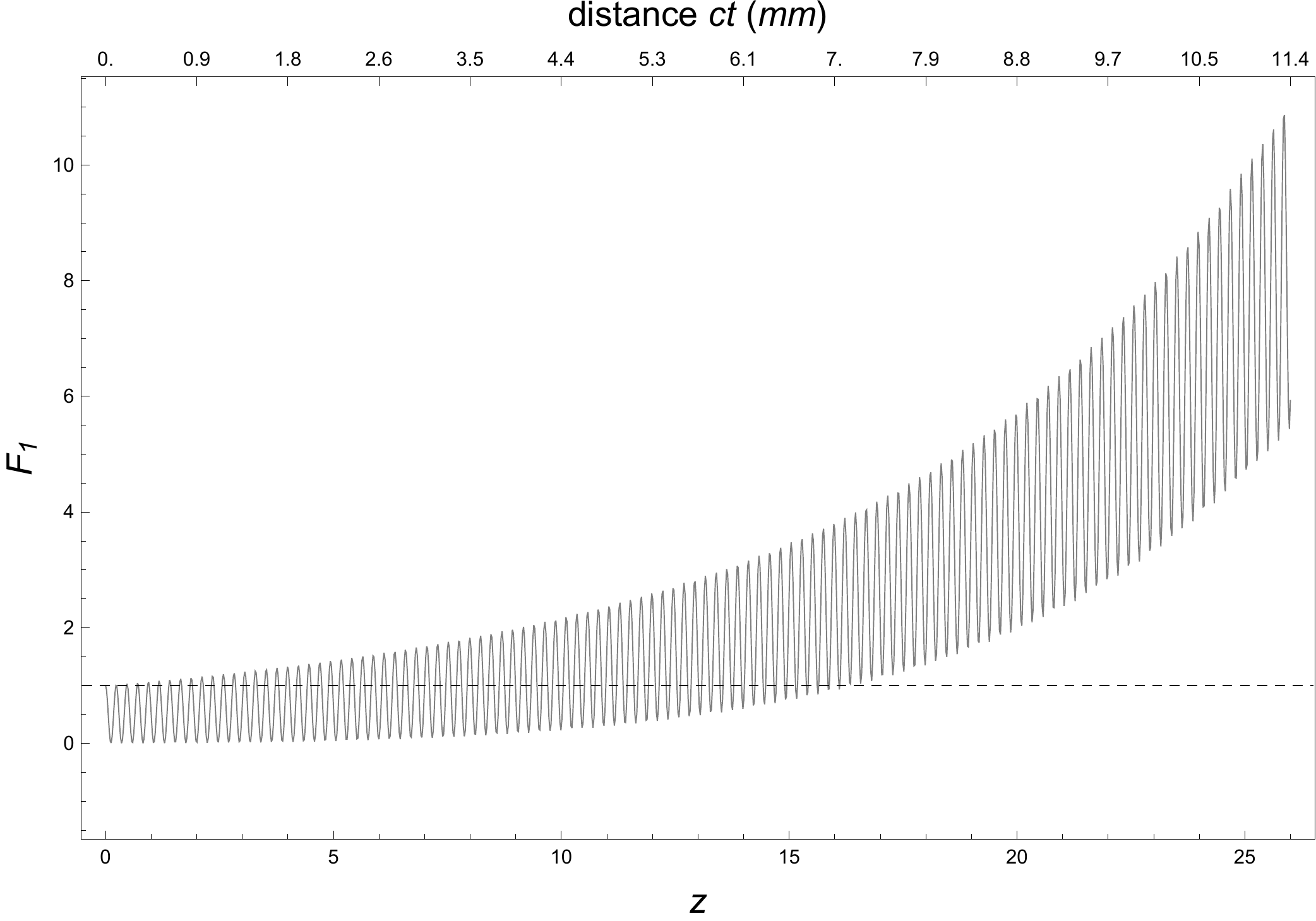} & \includegraphics[width=7.9cm]{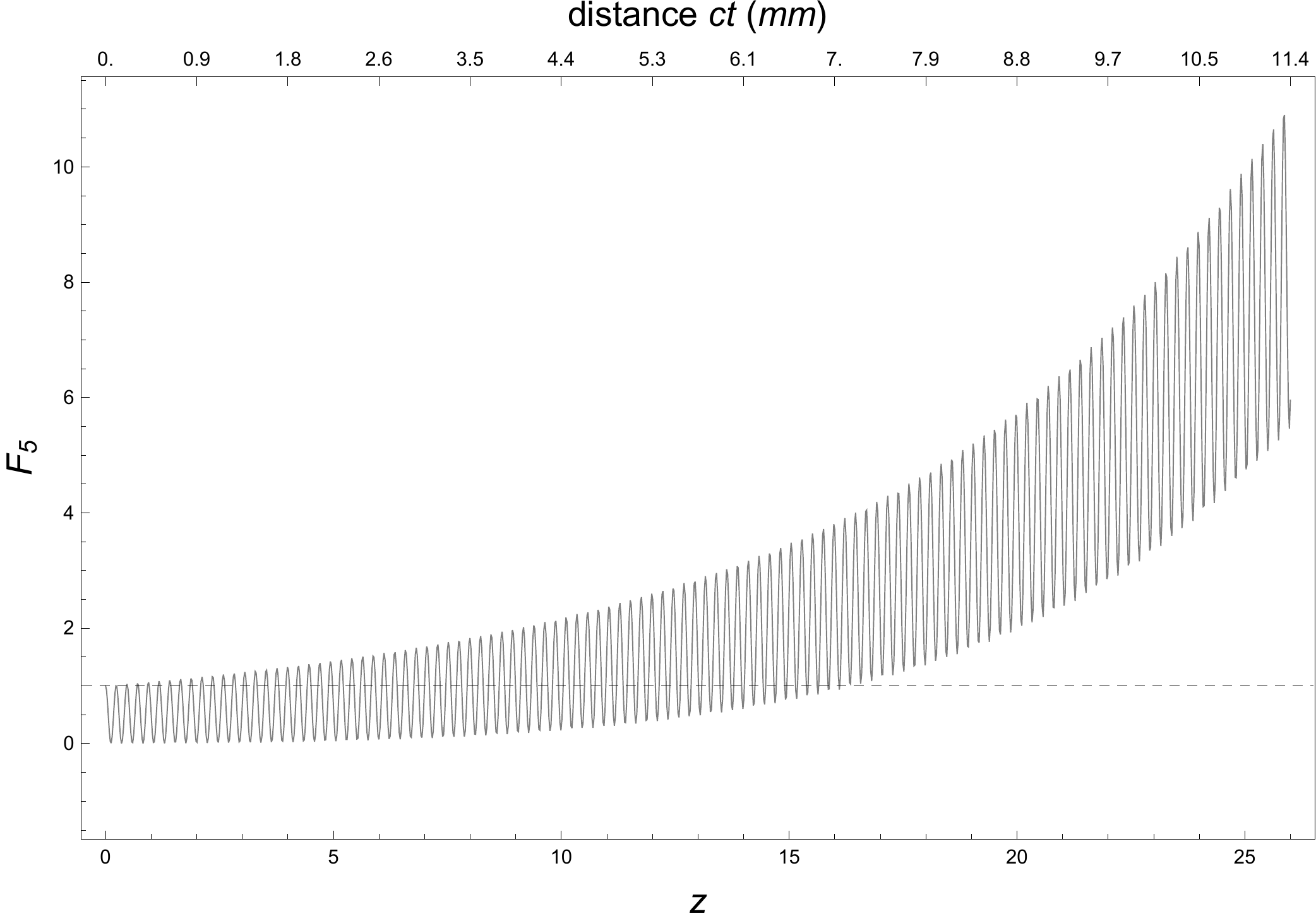}
	\end{tabular}
\caption{\protect\label{fig:F1_and_F5_Bs} Functions $\textrm{F}_{1,\,
    5} (x,\, r,\,\zeta,\,\lambda)$ for $B_s$-mesons. The top axis
  corresponds to $c \,t$ (in mm), the bottom axis, to the time in units
  of lifetime $z = (\Gamma_H + \Gamma_L)\, t /2 = \Gamma\, t =
  t/\tau_{B_s}$, where time $t$ is calculated in the $B_s$-meson rest
  frame. The plots are produced for $r = 1.004$ and $\zeta = 185^{\textrm{o}}$.
}
\end{figure}

If we introduce function 
$$
\textrm{F}_{1} (x,\, r,\,\zeta,\,\lambda)\, =\,  
     \left | 
              g_+(t)\, -\, \frac{1}{2}\,\left ( \frac{q}{p} + \frac{p}{q}\right )\, g_-(t)
     \right |^2\, |g_+ (t)|^2\,\, e^{2 \Gamma\, t},
$$
then inequality (\ref{timeM1M}) transforms into (\ref{F_N}). A plot of the
function $\textrm{F}_{1} (x,\, r,\,\zeta,\,\lambda)$ is shown in
Figure~\ref{fig:F1_and_F5_Bs} left. That taking into
account the oscillations, the violation of inequality
(\ref{timeM1M}) holds for 
$z = 1/\tau_{B_s} \ge 17$, which agrees with a na\"ive
estimate obtained from the simplified inequality
(\ref{timeM1M-simple}). One can also obtain inequality
(\ref{timeM1M}) by choosing events $\mathcal{S}_1 (t_0) $ and
$\mathcal{S}_2 (t) $ in the form $\mathcal{S}_1 (t_0) = \{M^{(2)}_1, \bar
M^{(1)}\}$, $\mathcal{S}_2 (t) = \{M^{(2)}_1, \bar M^{(1)}\}$. 
The event $\mathcal{S}_3(t_0)$ remains the same.

The explicit form of the functions $\textrm{F}_{N} (x,\,
r,\,\zeta,\,\lambda)$ is shown in Appendix \ref{sec:DD}.
The inequality (\ref{F_N}) in systems of neutral $B_s$-mesons is
violated not only while choosing events to which the function
$\textrm{F}_{1}$, corresponds, but also for sets of events, to which
correspond the functions $\textrm{F}_{2}$ and $\textrm{F}_{5}$. The
correspondence between sets of events and the functions is given in
Table~\ref{table:conditions}. The behaviour of the function
$\textrm{F}_{5}$ for $B_s$-mesons is shown in
Figure~\ref{fig:F1_and_F5_Bs} right. The dependence of functions
$\textrm{F}_{1}$ and $\textrm{F}_{5}$ on $z$ is almost identical  due
to the small magnitude of $CP$--violation effects. Unfortunately a test of
the violation of the inequality (\ref{F_N}) for $z \ge 17$ requires
large statistics due to the exponential character of $B_s$--meson decays.

\begin{figure}[bt]
	\begin{tabular}{cc}
         \includegraphics[width=7.9cm]{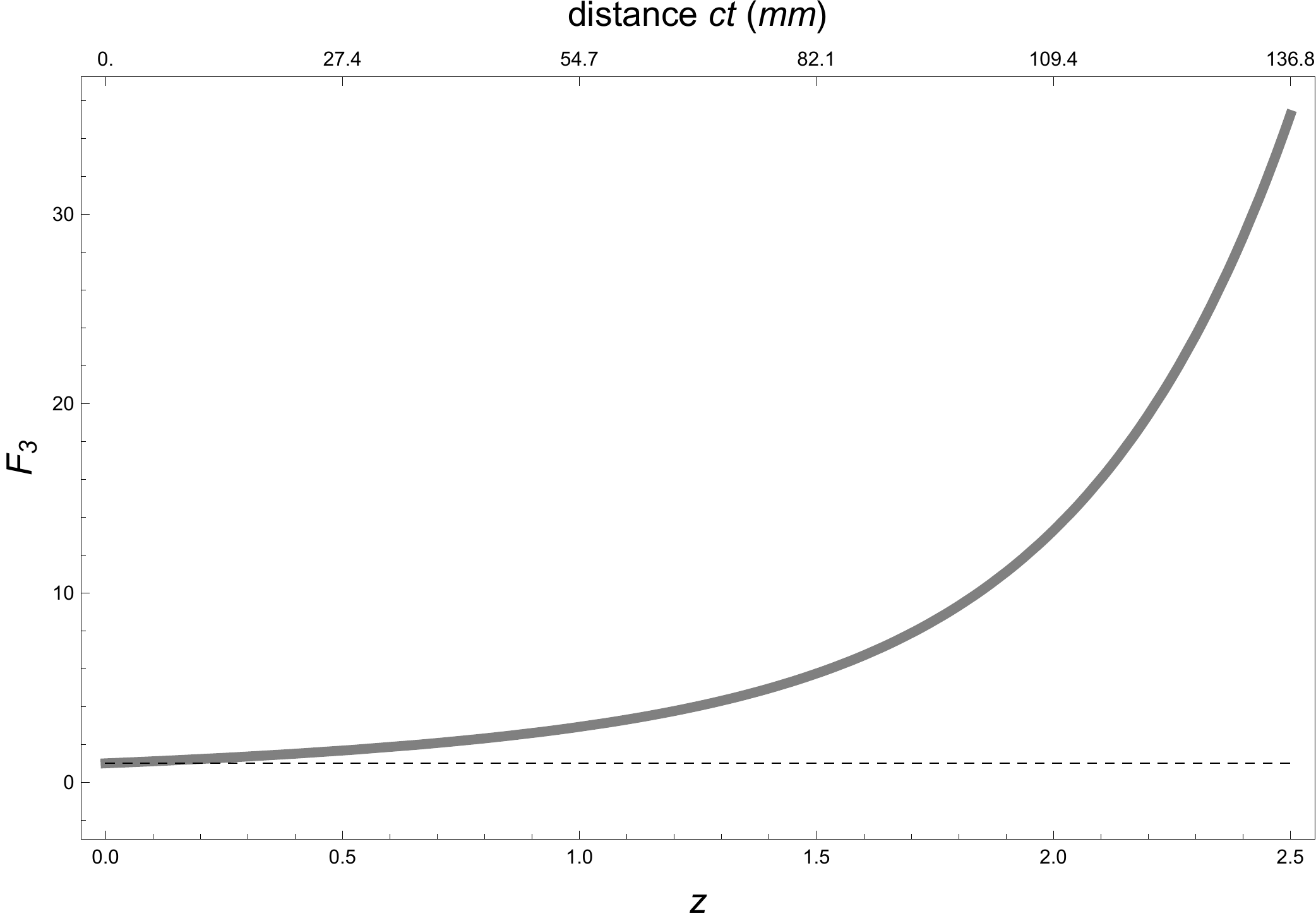} & \includegraphics[width=7.9cm]{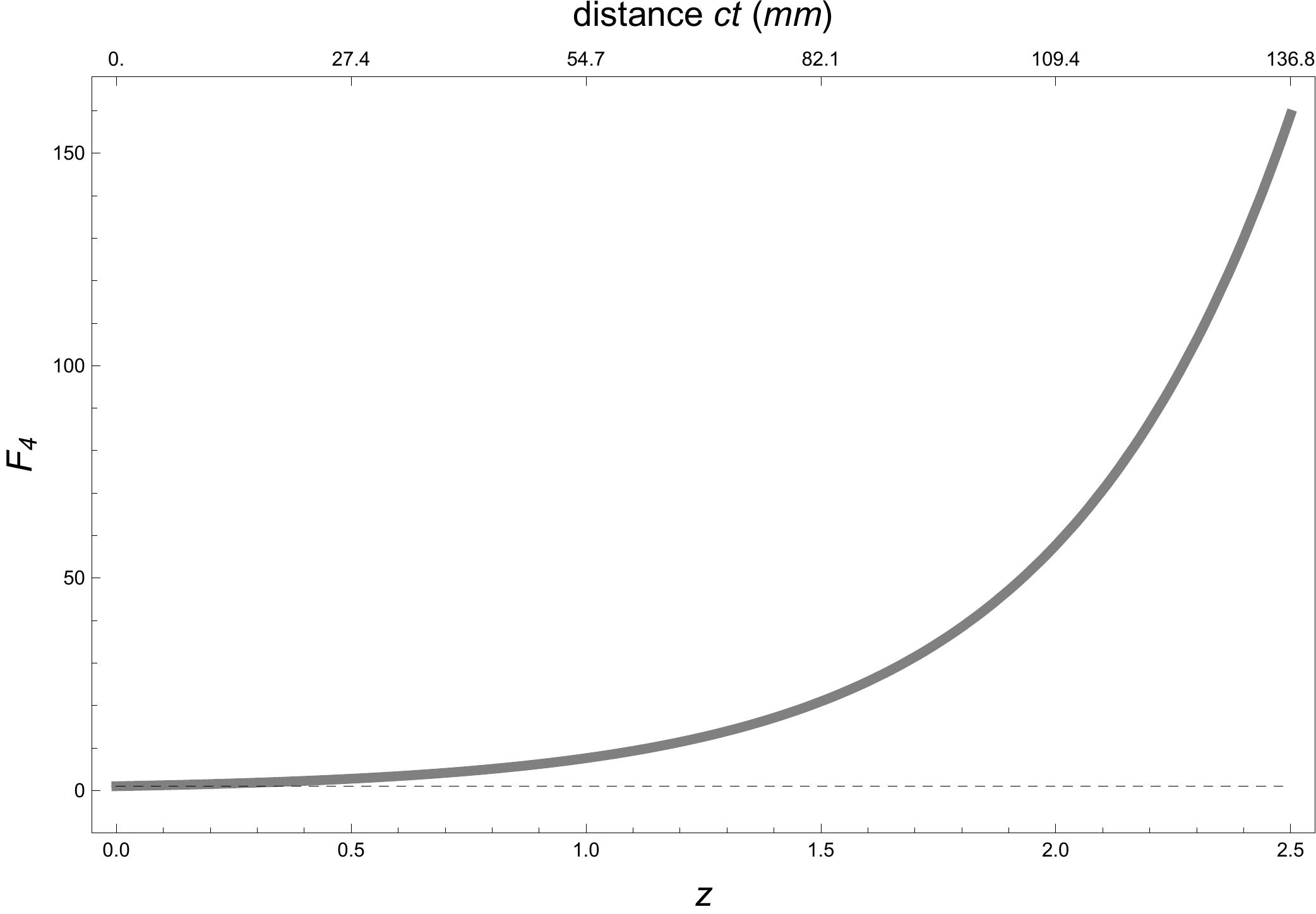}
	\end{tabular}
\caption{\protect\label{fig:F3_and_F4_K} Functions $\textrm{F}_{3,\,
    4} (x,\, r,\,\zeta,\,\lambda)$ for neutral $K$-mesons. The top axis
  corresponds to $c \,t$ (in mm), the bottom axis, to the time in units
  of lifetime $z = (\Gamma_S + \Gamma_L)\, t /2 = \Gamma\, t =
  t/\tau_{K}$, where $t$ is calculated in the $K$--meson rest
  frame. The plots are produced for $r = 0.997$ and $\zeta = -0.18^{\textrm{o}}$.
}
\end{figure}

Now consider systems of neutral kaons. According to
Table~\ref{table:conditions}, for $K$--mesons the inequality
(\ref{F_N}) is violated when choosing sets of events which lead to
the functions $\textrm{F}_{3}$, $\textrm{F}_{4}$, and
$\textrm{F}_{5}$. As one can see from Figure~\ref{fig:F3_and_F4_K}, a
significant violation of inequality (\ref{F_N}) holds for $z
\sim 1$, which makes the systems of neutral kaons good candidates for
an experimental test of the hypothesis of realism.

\begin{figure}[bt]
	\begin{tabular}{cc}
         \includegraphics[width=7.9cm]{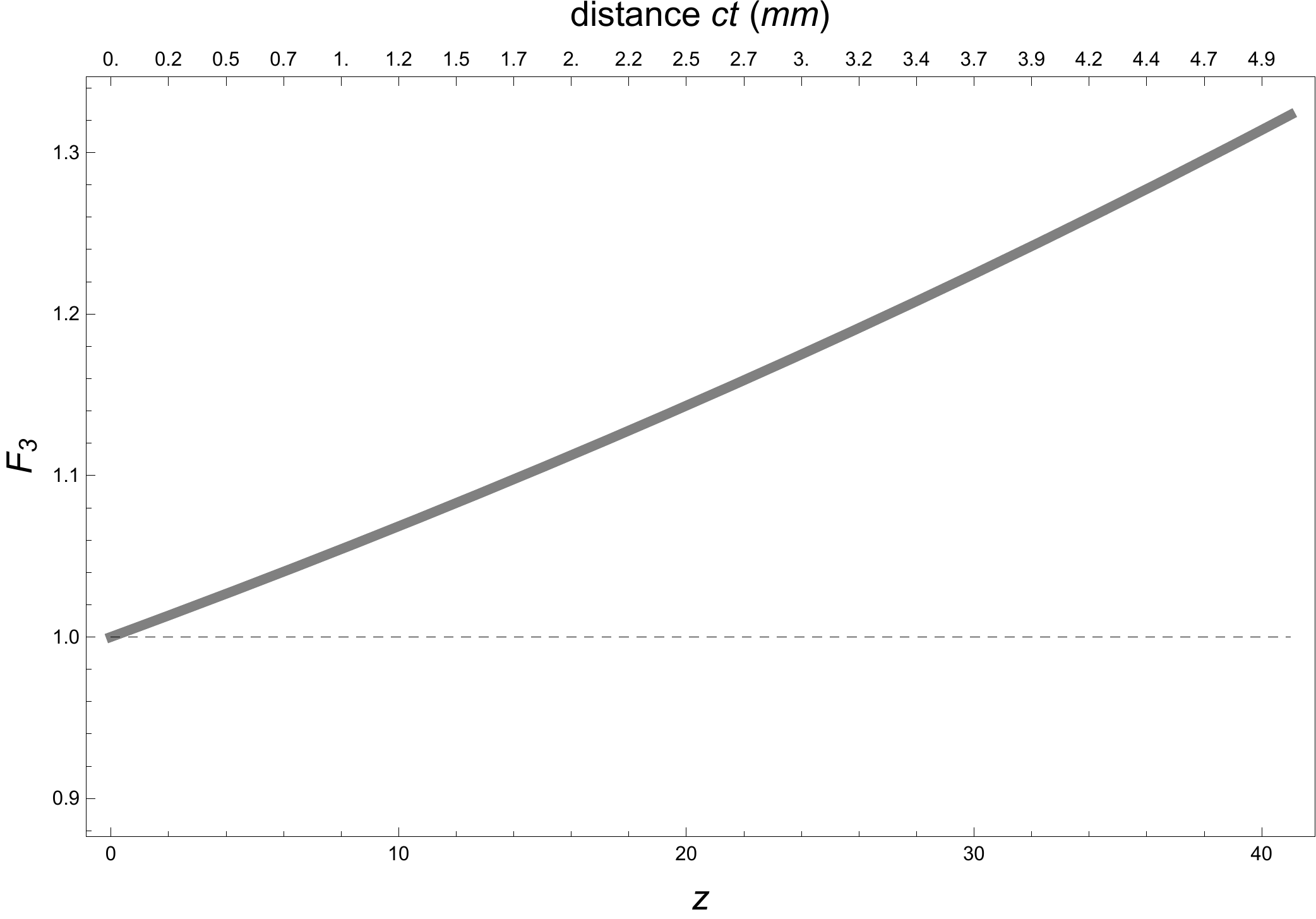} & \includegraphics[width=7.9cm]{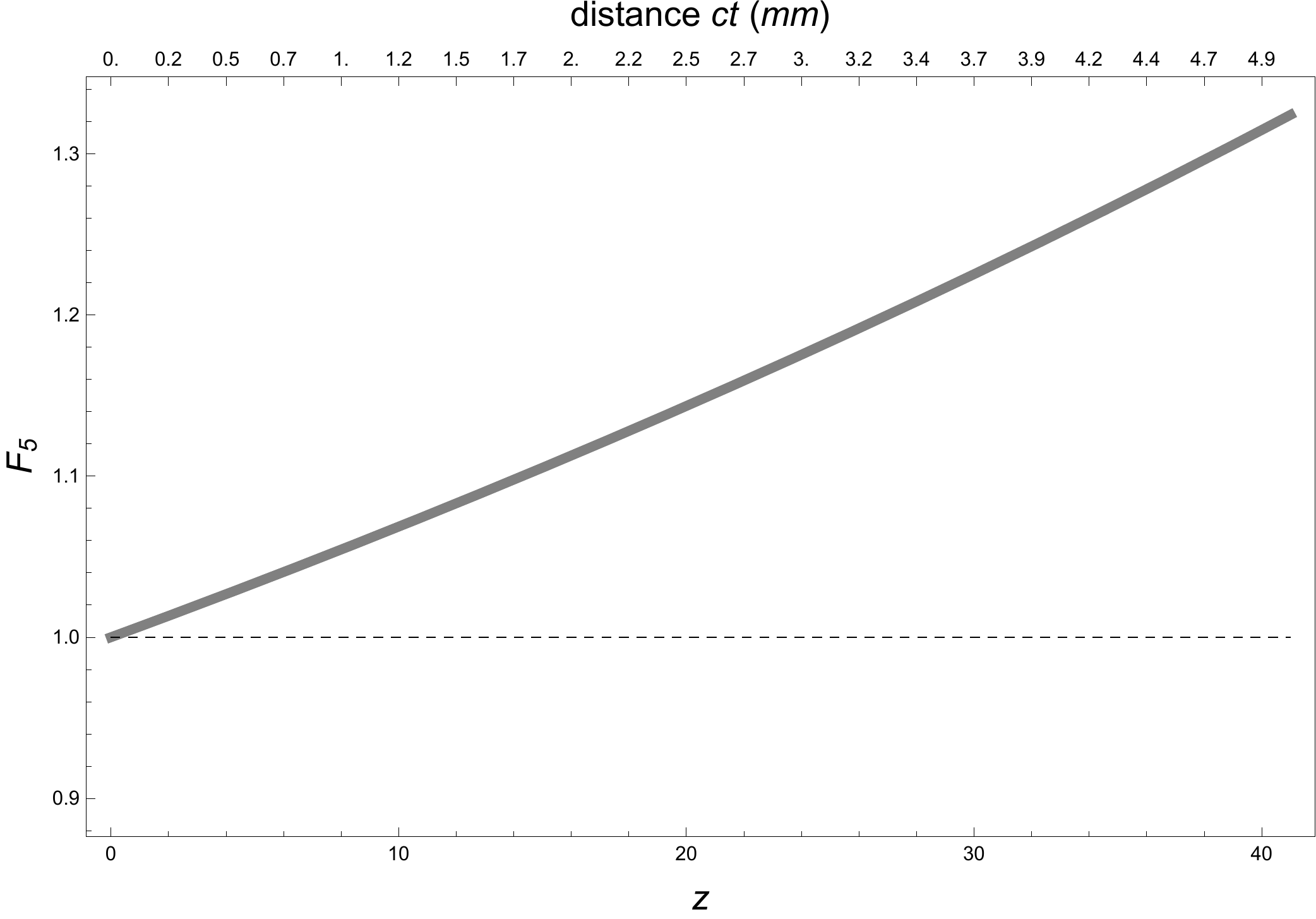}
	\end{tabular}
\caption{\protect\label{fig:F3_and_F5_D} Functions $\textrm{F}_{3,\,
    5} (x,\, r,\,\zeta,\,\lambda)$ for neutral $D$-mesons. The top
  axis corresponds to $c \,t$ (in mm), the bottom axis, to the time
  in units of lifetime $z = (\Gamma_H + \Gamma_L)\, t /2 = \Gamma\, t
  = t/\tau_{D}$, where $t$ is calculated in $D$--meson rest
  frame. The plots are produced for $r = 1.1$ and $\zeta = -10^{\textrm{o}}$.
}
\end{figure}

In the case of entangled states of $D^0\bar D^0$--mesons, the following
functions lead to violation of inequality (\ref{F_N}):
$\textrm{F}_{3}$, $\textrm{F}_{4}$, and $\textrm{F}_{5}$, which were
already considered for entangled kaons. This happens due to the fact
that for neutral $K$-- and $D$--mesons, the real part of the relation
$\displaystyle\frac{q}{p}$ is close to $+1$. From
Figure~\ref{fig:F3_and_F5_D}, for $D$--mesons even with $z \sim 40$,
the functions  $\textrm{F}_{3}$ and $\textrm{F}_{5}$ remain almost
linear. Note that  the corresponding functions for the
$K$--mesons demonstrate exponential growth already for $z \ge 1$ 
(see Figure~\ref{fig:F3_and_F4_K}). The difference in the behaviour
of the functions $\textrm{F}_{3,\, 4,\, 5} (x,\, r,\,\zeta,\,\lambda)$
for $K$-- and $D$--mesons is stipulated by the value of the relation
$|\Delta \Gamma | / \Gamma$, which sets the scale of the magnitude of
the functions. For $K$--mesons, $\displaystyle \left (\frac{|\Delta
    \Gamma |}{\Gamma} \right )_K \approx 2$, while for $D$--mesons
that parameter is lower by almost two orders of magnitude, equal
to $\displaystyle \left (\frac{|\Delta \Gamma |}{\Gamma} \right )_D
\approx 10^{-2}$. For $B_s$--mesons, $\displaystyle \left
  (\frac{|\Delta \Gamma |}{\Gamma} \right )_{B_s} \approx 0.13$,
and the value of $z\sim 15$ when the functions $\textrm{F}_{1,\, 2,\, 5}
(x,\, r,\,\zeta,\,\lambda)$ become exponential. Hence they are an
intermediate state between the values of $z$ for $K$-- and $D$--mesons.

The analysis described above shows that from the experimental point of
view, violation of inequality (\ref{F_N}) is more suitable to
observe in systems of entangled $K$- and $D$-mesons. Due to 
oscillations, for $B_s$--mesons inequality (\ref{F_N}) is
violated for $z \ge 17$ and its observation requires quite large statistics.

\newpage
\section{Conclusion}

Using the notion of conditional probability in the framework of
Kolmogorov's axiomatics and Bayes' theorem, we obtained the static
equality (\ref{bayes-static-bell}) and the time-dependent inequality
(\ref{bayes-dinamic-bell}) which allow experimental demonstration of the
unsoundness of the hypothesis of realism for quantum systems.  %

The structure of time-dependent inequality (\ref{bayes-dinamic-bell})
gives a principal possibility to avoid the contextuality
loophole which is still open in contemporary experiments that test
Bell's, Wigner's and the Leggett-Garg inequalities.

The possibility to experimentally test the violation of formulae
(\ref{bayes-static-bell}) and (\ref{bayes-dinamic-bell}) is studied
with two examples: the behaviour of correlated spins in a constant
and homogeneous magnetic field; and the behaviour of pairs of correlated
pseudoscalar mesons. Some factors that can prevent such tests at the
LHC experiments, Belle~II, and $\phi$-factories are considered.

\section{Acknowledgements}
The authors would like to express our deep gratitude to
Dr.~A.~Grinbaum (CEA-Saclay, France), Dr.~A.~A.Dzyuba (PNPI, Russia),
Dr.~P.~R.~Sharapova (Paderborn University, Germany), V.~P.~Sotnikov
(Albert Ludwig University of Freiburg, Germany), and A.~V.~Danilina
(MSU, Russia) for multiple helpful
discussions which improved the paper significantly. We would like to
thank Prof. S.~Seidel (University of New Mexico, USA) for help with
preparation of the paper. Individually we would like to thank
C. Aleister (Thoiry, France) for creating a warm and friendly working
atmosphere for discussions between the authors.

The work was supported by grant 16-12-10280 of the Russian Science
Foundation. One of the authors (N.~Nikitin) expresses his gratitude for
this support.

\newpage
\appendix

\section{Correlated spins in an external magnetic field. Main
  formulae}
\label{sec:A}

At the initial time $t_0 = 0$, a pseudoscalar particle at rest
decays into a positron (index ``$1$'') and an electron (index ``$2$''). If
such a decay is described by Hamiltonians (\ref{Heff_for_PS2ff}), then
for $t_0=0$ the $e^+ e^-$--pair is in Bell state $\ket{\Psi^-}$ with zero
full spin (\ref{psi-_t0}). 

Choose spatial direction 
$
\vec n = \left ( \sin\theta_n \cos\varphi_n, \, \sin\theta_n \sin\varphi_n,\, \cos\theta_n \right )
$. At time $t_0=0$, the state vectors of the positron and the electron, related
to the spin projections $\pm 1/2$ onto axis $\vec n$, are:
\begin{equation}
\label{wf-t0}
\ket{\frac{1}{2},\, n_+^{(i)}} = \left (
\begin{array}{c}
     \cos ( \nicefrac{\theta_n}{2} ) \, e^{- i \nicefrac{\varphi_n}{2}}\\
     \sin  ( \nicefrac{\theta_n}{2} ) \, e^{i \nicefrac{\varphi_n}{2}}
\end{array}
\right )\qquad \textrm{and}\qquad
\ket{\frac{1}{2},\, n_-^{(i)}} = \left (
\begin{array}{r}
     -\,\sin  (\nicefrac{\theta_n}{2}) \, e^{- i \nicefrac{\varphi_n}{2}}\\
     \cos  (\nicefrac{\theta_n}{2}) \, e^{i \nicefrac{\varphi_n}{2}}
\end{array}
\right ),
\end{equation}
where $i = \{1,\, 2\}$. Then
\begin{eqnarray}
\label{correlation-t=0}
\ket{\Psi^- (t_0)} &=& \frac{1}{\sqrt{2}}
\left [
\left (
\begin{array}{c}
     \cos\, (\nicefrac{\theta_n}{2})\, e^{- i\,\nicefrac{\varphi_n}{2}} \\
     \sin\,  (\nicefrac{\theta_n}{2})\, e^{i\,\nicefrac{\varphi_n}{2}} 
\end{array}
\right )^{(2)}
\left (
\begin{array}{c}
      - \sin\, (\nicefrac{\theta_n}{2})\, e^{- i\,\nicefrac{\varphi_n}{2}} \\
        \cos\, (\nicefrac{\theta_n}{2})\, e^{i\,\nicefrac{\varphi_n}{2}}  
\end{array}
\right )^{(1)} -  \right . \\
&-& \left .
\left (
\begin{array}{c}
      - \sin\, (\nicefrac{\theta_n}{2})\, e^{- i \nicefrac{\varphi_n}{2}} \\
        \cos\, (\nicefrac{\theta_n}{2})\, e^{i \nicefrac{\varphi_n}{2}}  
\end{array}
\right )^{(2)}
\left (
\begin{array}{c}
      \cos\, (\nicefrac{\theta_n}{2})\, e^{- i \nicefrac{\varphi_n}{2}} \\
      \sin\,  (\nicefrac{\theta_n}{2})\, e^{i \nicefrac{\varphi_n}{2}} 
\end{array}
\right )^{(1)}
\right ]. \nonumber
\end{eqnarray}
To illustrate the violation of relations
(\ref{bayes-static-bell}) and (\ref{neravenstvobayesa}) in NRQM, it is
enough to measure fermionic spin projections onto two or three
non-parallel directions lying in plane $(x,z)$ (we use a standard
rectangular coordinate system $(x,y,z)$). We map the unit
vectors $\vec a$, $\vec b$, and $\vec c$ to these directions. In this
case 
$
\varphi_a = \varphi_b = \varphi_c = 0
$. 

Now put this spin singlet $e^+ e^-$--state into a constant,
homogeneous magnetic field with strength $\vec{\mathcal{H}}$ directed
along the $y$-axis. Require the electron and positron to propagate
strictly along $y$. This requirement avoids unnecessary
complications related to the rotation of charged particles in the magnetic
field.   

The spins of the electron and positron will begin to precess around the 
$y$-axis. Given initial condition (\ref{wf-t0}), the state vectors
of the electron which describe its spin projections onto $\vec n$ at
an arbitraty moment of time may be written as 
\begin{eqnarray}
\label{wf-t-e-}
&& 
\ket{\psi^{(2)}_{n_+}(t)} = \left (
\begin{array}{c}
     \displaystyle \cos \nicefrac{\theta_n}{2}\, \cos ( \omega t)\, e^{- i \varphi_n / 2} -  \sin \nicefrac{\theta_n}{2}\, \sin ( \omega t)\, e^{i \varphi_n / 2}\\
     \displaystyle \cos \nicefrac{\theta_n}{2}\, \sin ( \omega t)\, e^{- i \varphi_n / 2} +  \sin \nicefrac{\theta_n}{2}\, \cos ( \omega t)\, e^{i \varphi_n / 2}
\end{array}
\right )^{(2)}\qquad \textrm{and} \\
&&
\ket{\psi^{(2)}_{n_-}(t)} =
\left (
\begin{array}{r}
       \displaystyle -\,\sin\nicefrac{\theta_n}{2}\, \cos ( \omega t)\, e^{-i \varphi_n / 2} -\,\cos \nicefrac{\theta_n}{2}\, \sin ( \omega t)\, e^{i \varphi_n / 2} \\
       \displaystyle -\,\sin\nicefrac{\theta_n}{2}\, \sin ( \omega t)\, e^{-i \varphi_n / 2} +\,\cos \nicefrac{\theta_n}{2}\, \cos ( \omega t)\, e^{i \varphi_n / 2}
\end{array}
\right )^{(2)}. \nonumber
\end{eqnarray}
For the positron the analogous state vectors are
\begin{eqnarray}
\label{wf-t-e+}
&&
\ket{\psi^{(1)}_{n_+}(t)} = \left (
\begin{array}{r}
     \displaystyle \cos \nicefrac{\theta_n}{2}\, \cos ( \omega t)\, e^{- i \varphi_n / 2} + \sin \nicefrac{\theta_n}{2}\, \sin ( \omega t)\, e^{i \varphi_n / 2}\\
     \displaystyle -\,\cos \nicefrac{\theta_n}{2}\, \sin ( \omega t)\, e^{- i \varphi_n / 2} +  \sin \nicefrac{\theta_n}{2}\, \cos ( \omega t)\, e^{i \varphi_n / 2}    
\end{array}
\right )^{(1)}\qquad \textrm{and}\\
&&
\ket{\psi^{(1)}_{n_-}(t)} =
\left (
\begin{array}{c}
        \displaystyle -\,\sin\nicefrac{\theta_n}{2}\, \cos ( \omega t)\, e^{-i \varphi_n / 2} +\,\cos \nicefrac{\theta_n}{2}\, \sin ( \omega t)\, e^{i \varphi_n / 2} \\
       \displaystyle      \sin\nicefrac{\theta_n}{2}\, \sin ( \omega t)\, e^{-i \varphi_n / 2} +\,\cos \nicefrac{\theta_n}{2}\, \cos ( \omega t)\, e^{i \varphi_n / 2}
      \end{array}
\right )^{(1)},\nonumber
\end{eqnarray}
where $\displaystyle\omega = \frac{|e|\, \mathcal{H}}{2 m_e c}$,
the Larmor frequency of a fermion.

Using initial condition (\ref{correlation-t=0}) and the explicit form
of the wave functions of the electron (\ref{wf-t-e-}) and positron
(\ref{wf-t-e+}) in the magnetic field, we obtain for the spin
wave function of the $e^+e^-$-pair for arbitrary time $t \ge t_0$:
\begin{eqnarray}
\label{psi-_t}
\ket{\Psi^- (t)}\, =\,\frac{1}{\sqrt{2}}\, 
\Big  (
\ket{\psi^{(2)}_{{\bf n}_+}(t)}\, \ket{\psi^{(1)}_{{\bf n}_-}(t)}\, -\, \ket{\psi^{(2)}_{{\bf n}_-}(t)}\, \ket{\psi^{(1)}_{{\bf n}_+}(t)}
\Big ).
\end{eqnarray}

\section{Oscillations of neutral pseudoscalar mesons. Main formulae}
\label{sec:B}

The definitions used in this Section are analogous to those from
\cite{Nikitin:2015bca} and \cite{Nikitin:2015mna}.

In contrast to spin states for which one can choose an infinite number of
spatial directions for neutral pseudoscalar mesons $M =
\{ K,\, D,\, B_q \}$, where $q = \{ d,\, s\}$,  there are only three
fixed ``directions'' with corresponding non-commuting ``projectors''. 

As a first ``direction'' let us choose the flavour of the pseudoscalar
meson. For $D$--mesons, consider projections
onto states $\ket{D} = \ket{c \bar u}$ and $\ket{\bar D} = \ket{\bar c
  u}$.  Operators of charge ($\hat C$) and space ($\hat P$)
conjugation act on the flavour states as follows: 
\begin{eqnarray}
\hat C \hat P\,\ket{M}\, =\, e^{i\alpha} \ket{\bar M}\quad\textrm{and}\quad\hat C \hat P\,\ket{\bar M}\, =\, e^{-i \alpha} \ket{M}, \nonumber
\end{eqnarray}
where $\alpha$ is an arbitrary real phase. This phase should not
appear in any experimentally-testable relations. States $\ket{M}$ and
$\ket{\bar M}$ are orthogonal to each other.

A second ``direction'' is specified by the states with definite values of $CP$-parity: 
\begin{eqnarray}
\hat C \hat P\,\ket{M_1}\, =\, +\,\ket{M_1},\qquad \hat C \hat P\,\ket{M_2}\, =\, -\,\ket{M_2}, \nonumber
\end{eqnarray}
which can be written using the states $\ket{M}$ and $\ket{\bar M}$ as
\begin{eqnarray}
\ket{M_1} = \frac{1}{\sqrt{2}}\,\left ( \ket{M} + e^{i\alpha} \ket{\bar M}\right ), \quad 
\ket{M_2} = \frac{1}{\sqrt{2}}\,\left ( \ket{M} - e^{i\alpha} \ket{\bar M}\right ). \nonumber
\end{eqnarray}
Note that $\bracket{M_1}{M_2} = 0$.

A third ``direction'' corresponds to the states with definite lifetimes
and masses. In terms of $\ket{M}$ and $\ket{\bar M}$, projections onto
this ``direction'' may be written as
\begin{eqnarray}
\ket{M_L} = p\left ( \ket{M} + e^{i \alpha}\,\frac{q}{p}\, \ket{\bar M}\right ) \quad\textrm{and}\quad
\ket{M_H} = p\left (\ket{M} - e^{i \alpha}\,\frac{q}{p}\, \ket{\bar M} \right ) . \nonumber
\end{eqnarray}
Using the normalization condition, we find the relation for complex
coefficients $p$ and $q$:
\begin{eqnarray}
\label{pq-normirovka}
\bracket{M_L}{M_L}=\bracket{M_H}{M_H}=|p|^2 + |q|^2 = 1. 
\end{eqnarray}
It can be shown that $\bracket{M_L}{M_H} = |p|^2 - |q|^2 \ne 0$.

To automatically satisfy the normalization condition
(\ref{pq-normirovka}) we introduce a
new variable $\beta$: 
\begin{eqnarray}
| p | = \cos\beta ;\qquad |q|=\sin\beta \qquad \textrm{and}\qquad \frac{q}{p} = \tg \beta \, e^{i \zeta} \equiv r  e^{i \zeta}. \nonumber
\end{eqnarray}
From the definition it follows that $\beta \in \left [ 0,\, \pi/2 \right ]$. 

Taking into account $CPT$--invariance, the states $\ket{M_L}$ and
$\ket{M_H}$ are eigenvectors of the Hamiltonian 
$$
\hat H =  
\left (
\begin{array}{lr}
    \mathcal{H}              &  H_{12}\, e^{-i \alpha}\\
    H_{21}\, e^{i \alpha} &   \mathcal{H}
\end{array}
\right )\, =\,
\left (
\begin{array}{lr}
    m - \nicefrac{i}{2}\,\Gamma                                                                        &   \left ( m_{12} - \nicefrac{i}{2}\,\Gamma_{12} \right )\, e^{-i \alpha}\\
    \left ( m_{12}^* - \nicefrac{i}{2}\,\Gamma_{12}^* \right )\, e^{i \alpha} &   m - \nicefrac{i}{2}\,\Gamma
\end{array}
\right ),
$$
with eigenvalues 
\begin{eqnarray}
&& E_L  = m_L - \nicefrac{i}{2}\, \Gamma_L =   \mathcal{H} - \sqrt{H_{12} H_{21}} =   \mathcal{H} + \nicefrac{q}{p}\, H_{12}\quad \textrm{and} \nonumber \\
&& E_H = m_H - \nicefrac{i}{2}\, \Gamma_H  =  \mathcal{H} + \sqrt{H_{12} H_{21}} =   \mathcal{H} - \nicefrac{q}{p}\, H_{12} \nonumber
\end{eqnarray}
accordingly. Finally we define parameters
\begin{eqnarray}
&&  \Delta M = M_H - M_L = -\, 2\, \textrm{Re}\,\left ( \frac{q}{p}\, H_{12}\right ), \nonumber \\
&& \Delta \Gamma = \Gamma_H - \Gamma_L = 4\, \textrm{Im}\, \left ( \frac{q}{p}\, H_{12}\right ). \nonumber
\end{eqnarray}
Please note that the definition of $\Delta \Gamma$ here
differs by a sign from the definition of $\Delta \Gamma$
in \cite{Beringer:1900zz}. Experimental values of the parameters of
$CP$--violation are shown in Table~\ref{table:parameters}.

\begin{table}[tb]
\hfill
\bigskip
\begin{center}
\begin{tabular}{||c|c|r|r|c||}
\hline
\hline
Meson & $\Delta\Gamma$, MeV & $\Delta M$, MeV & $\tg \beta = \left|\nicefrac{q}{p}\right |^{\textrm{exp}}_M$ & $\lambda$\\ 
\hline\hline 
$B^{0}_{s}$ & $-\, 6.0\times 10^{-11}$ & $1.2\times 10^{-8}$& $1.0039\pm 0.0021$                  & $- 0.2 \times 10^{3}$ \\ 
$K^{0}$        &  $-\, 7.3\times 10^{-12}$ & $3.5\times 10^{-12}$ & $0.99668\pm 0.00004$          & $- 4.8 \times 10^{-1}$ \\ 
$D^{0}$        & $ -\, 2.1\times 10^{-11}$ & $ -\, 6.3\times 10^{-12}$ & $0.92^{+0.12}_{-0.09}$  & $0.3$ \\
\hline
\hline
\end{tabular} 
\end{center}
\caption{Experimentally found parameters of oscillations and
  $CP$-violation for systems of neutral pseudoscalar mesons. The table
  is modelled on one in \protect\cite{Beringer:1900zz}. The minus sign in
  the numerical values of $\Delta\Gamma$ reflects the difference in
  definitions between the current paper and
  \protect\cite{Beringer:1900zz}. Dimensionless variable $\lambda =
  \nicefrac{\Delta M}{\Delta\Gamma}$).}
\label{table:parameters}
\end{table}

Decay of a neutral vector meson with quantum numbers $J^{P\, C} = 1^{-\,
  -}$  into a pair of pseudoscalar mesons (experiments mostly deal with
the decays $\phi (1020) \to K \bar K$, $\Upsilon (4S) \to 
B_d \bar B_d$, and $\Upsilon (5S) \to B_s \bar B_s$) produces a state
of an $M \bar M$-pair at the time $t_0 = 0$ which is described by the Bell
state vector 
\begin{eqnarray}
\label{correlationBbarB-t=0}
\ket{\Psi^- (t_0)} &=& \frac{1}{\sqrt{2}}
\left (\ket{M^{(2)}} \ket{\bar M^{(1)}}\, -\,\ket{\bar M^{(2)}} \ket{M^{(1)}}\right )\, = \nonumber \\
&=& \frac{e^{- i \alpha}}{\sqrt{2}}
\left (\ket{M_2^{(2)}} \ket{M_1^{(1)}}\, -\,\ket{M_1^{(2)}} \ket{M_2^{(1)}}\right )\, = \\
&=& \frac{1}{2\,\sqrt{2}\, p\, q}
\left (\ket{M_H^{(2)}} \ket{M_L^{(1)}}\, -\,\ket{M_L^{(2)}} \ket{M_H^{(1)}}\right ). \nonumber 
\end{eqnarray}
This state vector is fully analogous to the state vector
(\ref{correlation-t=0}), entangled in the spin space
\cite{Uchiyama:1996va,Bertlmann:2001sk,Bertlmann:2001ea}).

The evolution of the state vectors $\ket{M_L}$ and $\ket{M_H}$ can be written as:
\begin{eqnarray}
\ket{M_L (t)} &=&  e^{- i E_L \,\Delta t}  \ket{M_L} = e^{- i m_L \,\Delta t - \Gamma_L\, \Delta t /2} \ket{M_L}, \\
\ket{M_H (t)} &=&  e^{- i E_H \,\Delta t}  \ket{M_H} = e^{- i m_H \,\Delta t - \Gamma_H\, \Delta t / 2} \ket{M_H}, \nonumber
\end{eqnarray}
where $\Delta t = t - t_0$.
From the above one can find the evolution of the states $\ket{M (t)}$ and $\ket{\bar M (t)}$:
\begin{eqnarray}
\left \{
\begin{array}{l}
\displaystyle \ket{M(t)} = g_+(\Delta t) \ket{M}\, -\, e^{i \alpha}\,\frac{q}{p}\, g_-(\Delta t) \ket{\bar M} \\
\displaystyle \ket{\bar M (t)} = g_+(\Delta t) \ket{\bar M} - e^{-i \alpha}\,\frac{p}{q}\, g_-(\Delta t) \ket{M}
\end{array}
\right .  \nonumber
\end{eqnarray}
and the time dependence of the state vectors $\ket{M_1(t)}$ and $\ket{M_2(t)}$:
\begin{eqnarray}
\ket{M_1(t)} = \frac{1}{\sqrt{2}}
\left (
\left ( 
g_+(\Delta t) - \,\frac{p}{q}\, g_-(\Delta t)
\right ) \ket{M}\, +\,
e^{i \alpha}\,\left ( 
g_+(\Delta t)\,  -\,\frac{q}{p}\, g_-(\Delta t)
\right ) \ket{\bar M}
\right ),
\nonumber \\
\ket{M_2(t)} = \frac{1}{\sqrt{2}}
\left (
\left ( 
g_+(\Delta t) + \,\frac{p}{q}\, g_-(\Delta t)
\right ) \ket{M}\, -\,
e^{i \alpha}\,\left ( 
g_+(\Delta t)\,  +\,\frac{q}{p}\, g_-(\Delta t)
\right ) \ket{\bar M}
\right ),
\nonumber
\end{eqnarray}
where $\displaystyle g_{\pm}(\tau) = \frac{1}{2}\,\left ( e^{-i E_H
    \tau} \pm e^{- i E_L \tau}\right )$. Function $g_{\pm}(\tau)$
satisfies the conditions: 
\begin{eqnarray}
&& \left | g_{\pm}(\tau) \right |^2 = \frac{e^{-\Gamma \tau}}{2}\,
\left (
\ch \left ( \frac{\Delta\Gamma\, \tau}{2}\right ) \pm \cos \left ( \Delta M\, \tau\right ) \right ) \nonumber \\
&& g_+^* (\tau) g_-(\tau) = -\,\frac{e^{-\Gamma \tau}}{2}\,
\left (
\sh \left ( \frac{\Delta\Gamma\, \tau}{2}\right ) +  i\sin \left ( \Delta M\, \tau\right ) 
\right ),  \nonumber 
\end{eqnarray}
where $\Gamma = (\Gamma_H + \Gamma_L)/2$. Taking into account the
initial condition (\ref{correlationBbarB-t=0}), for the state vector
of the $M\bar M$-pair at an arbitrary time one can write:
\begin{eqnarray}
\label{correlationBbarB-t}
\ket{\Psi^- (t)} =  e^{-i \left ( m_H + m_L\right )\, \Delta t}\, e^{- \Gamma\, \Delta t}\,\ket{\Psi^- (t_0)}.
\end{eqnarray}
For $t_0 =0$ above, $\Delta t \equiv t$.

In systems of neutral pseudoscalar mesons, the magnitude of
$CP$--violation is small. If we neglect the $CP$--violation which
appears due to oscillations, then for $K$--mesons, 
$
\displaystyle \left (\frac{q}{p} \right )_K = \frac{1 - \epsilon}{1 + \epsilon} \approx 1
$;
so $\cos\zeta_K = 1$. For $B_q$--mesons the effective Hamiltonian of
the oscillations is proportional to $\left ( V_{tb} V_{tq}^*\right )^2$ \cite{Buras:2001ra}. Then
$$
\displaystyle \left (\frac{q}{p} \right )_{B_q} = -\,\frac{H_{21}}{\sqrt{H_{12}\, H_{21}}} \approx \, -\,
\left (\frac{V_{tb}^* V_{tq}}{\left | V_{tb}^* V_{tq} \right |} \right )^2 = -1,
$$ 
hence $\cos\zeta_{B_q} = -1$. For $D$--mesons, experimental data
from BaBar \cite{delAmoSanchez:2010xz} and Belle \cite{Abe:2007rd} are
in agreement with the assumption that $\cos\zeta_D = 1$, so $\cos\zeta =
\pm 1$ is a good approximation and the analysis of formulae
(\ref{bayes-static-bell}) and (\ref{bayes-dinamic-bell}) is much simplified.

\section{Oscillations of neutral pseudoscalar mesons. Transition
  probabilities}
\label{sec:D}

In this Section we collect the probabilities that are necessary for
a test of the static equality (\ref{bayes-static-bell}) and
time-dependent inequality (\ref{bayes-dinamic-bell}) in systems of
neutral pseudoscalar mesons.

In the framework of quantum theory using the normalization condition and
the initial condition (\ref{correlationBbarB-t=0}), the
following expressions for time-independent probabilities hold:
\begin{eqnarray}
\label{w-BbarB-I}
&& w(M_1^{(2)},\, \bar M^{(1)},\, t_0)\,=\,  \left |\bra{M_1^{(2)}}\bracket{\bar M^{(1)}}{\Psi^- (t_0)}\right |^2\, =\,\frac{1}{4}\,\equiv
\, \frac{1}{4}\,\left ( |p|^2 + |q|^2 \right );\nonumber \\
&& w(M_1^{(2)},\, M^{(1)},\, t_0)\,=\,  \left |\bra{M_1^{(2)}}\bracket{M^{(1)}}{\Psi^- (t_0)}\right |^2\, =\,\frac{1}{4}\,\equiv
\,\frac{1}{4}\,\left ( |p|^2 + |q|^2 \right );\nonumber \\
&& w(M_2^{(2)},\, \bar M^{(1)},\, t_0)\,=\,  \left |\bra{M_2^{(2)}}\bracket{\bar M^{(1)}}{\Psi^- (t_0)}\right |^2\, =\,\frac{1}{4}\,\equiv
\, \frac{1}{4}\,\left ( |p|^2 + |q|^2 \right );\nonumber \\
&& w(M_2^{(2)},\, M^{(1)},\, t_0)\,=\,  \left |\bra{M_2^{(2)}}\bracket{M^{(1)}}{\Psi^- (t_0)}\right |^2\, =\,\frac{1}{4}\,\equiv
\,\frac{1}{4}\,\left ( |p|^2 + |q|^2 \right );\\
&& w(M_1^{(2)},\, M_H^{(1)},\, t_0)\,=\, \left |\bra{M_1^{(2)}}\bracket{M^{(1)}_H}{\Psi^- (t_0)}\right |^2\, =\,
\frac{1}{4}\, \left |p + q\right |^2; \nonumber \\
&& w(M_2^{(2)},\, M_H^{(1)},\, t_0)\,=\, \left |\bra{M_2^{(2)}}\bracket{M^{(1)}_H}{\Psi^- (t_0)}\right |^2\, =\,
\frac{1}{4}\, \left | p - q\right |^2; \nonumber \\
&& w(M_1^{(2)},\, M_L^{(1)},\, t_0)\,=\, \left |\bra{M_1^{(2)}}\bracket{M^{(1)}_L}{\Psi^- (t_0)}\right |^2\, =\,
\frac{1}{4}\, \left |p - q\right |^2; \nonumber \\
&& w(M_2^{(2)},\, M_L^{(1)},\, t_0)\,=\, \left |\bra{M_2^{(2)}}\bracket{M^{(1)}_L}{\Psi^- (t_0)}\right |^2\, =\,
\frac{1}{4}\, \left | p + q\right |^2; \nonumber \\
&& w(M_H^{(2)},\, \bar M^{(1)},\, t_0)\,=\,\left |\bra{M_H^{(2)}}\bracket{\bar M^{(1)}}{\Psi^- (t_0)}\right |^2\, =\,
\frac{1}{2}\, \left | p \right |^2; \nonumber \\
&& w(M_H^{(2)},\, M^{(1)},\, t_0)\,=\,\left |\bra{M_H^{(2)}}\bracket{M^{(1)}}{\Psi^- (t_0)}\right |^2\, =\,
\frac{1}{2}\, \left | q \right |^2; \nonumber \\
&& w(\bar M^{(2)},\, M_L^{(1)},\, t_0)\,=\,\left |\bra{\bar M^{(2)}}\bracket{M_L^{(1)}}{\Psi^- (t_0)}\right |^2\, =\,
\frac{1}{2}\, \left | p \right |^2; \nonumber \\
&& w(M^{(2)},\, M_L^{(1)},\,  t_0)\,=\,\left |\bra{M^{(2)}}\bracket{M_L^{(1)}}{\Psi^- (t_0)}\right |^2\, =\,
\frac{1}{2}\, \left | q \right |^2. \nonumber 
\end{eqnarray}

In order to test the time-dependent inequality
(\ref{bayes-dinamic-bell}) for correlated $M \bar M$-pairs, the
following time-dependent probabilities are needed (for $t_0 = 0$):
\begin{eqnarray}
\label{w-BbarB-II}
&& w(M_1(0) \to M_1(t)) =  \left |\bracket{M_1(t)}{M_1} \right |^2= 
      \left | 
              g_+(t)\, -\, \frac{1}{2}\,\left ( \frac{q}{p} + \frac{p}{q}\right )\, g_-(t)
     \right |^2; \nonumber \\
&& w(M_2 (0) \to M_1 (t)) =  \left |\bracket{M_1(t)}{M_2} \right |^2=  \left | \frac{1}{2}\,\left ( \frac{q}{p} - \frac{p}{q}\right )\ g_- (t) \right |^2; \nonumber \\
&& w(M_2(0) \to M_2(t)) =  \left |\bracket{M_2(t)}{M_2} \right |^2= 
      \left | 
              g_+(t)\, +\, \frac{1}{2}\,\left ( \frac{q}{p} + \frac{p}{q}\right )\, g_-(t)
     \right |^2; \nonumber \\
&& w(M_1 (0) \to M_2 (t)) =  \left |\bracket{M_2(t)}{M_1} \right |^2=  \left | \frac{1}{2}\,\left ( \frac{q}{p} - \frac{p}{q}\right )\ g_- (t) \right |^2; \nonumber \\
&& w(\bar M (0) \to \bar M (t)) =  \left |\bracket{\bar M (t)}{\bar M} \right |^2 =  |g_+ (t)|^2; \nonumber \\
&& w(M (0) \to \bar M (t)) =  \left |\bracket{\bar M (t)}{M} \right |^2 = \left | \frac{p}{q}\, g_- (t) \right |^2; \\
&& w( M (0) \to M (t)) =  \left |\bracket{M (t)}{M} \right |^2 =  |g_+ (t)|^2; \nonumber \\
&& w(\bar M (0) \to M (t)) =  \left |\bracket{M (t)}{\bar M} \right |^2 = \left | \frac{q}{p}\, g_- (t) \right |^2; \nonumber \\
&&w(M_1^{(2)},\, \bar M^{(1)},\, t) = \left |\bra{M_1^{(2)}}\bracket{\bar M^{(1)}}{\Psi^- (t)}\right |^2\, =\,\frac{1}{4}\, e^{-2 \Gamma\, t}; \nonumber \\
&&w(M_1^{(2)},\, M^{(1)},\, t) = \left |\bra{M_1^{(2)}}\bracket{M^{(1)}}{\Psi^- (t)}\right |^2\, =\,\frac{1}{4}\, e^{-2 \Gamma\, t}; \nonumber \\
&&w(M_2^{(2)},\, \bar M^{(1)},\, t) = \left |\bra{M_2^{(2)}}\bracket{\bar M^{(1)}}{\Psi^- (t)}\right |^2\, =\,\frac{1}{4}\, e^{-2 \Gamma\, t}; \nonumber \\
&&w(M_2^{(2)},\, M^{(1)},\, t) = \left |\bra{M_2^{(2)}}\bracket{M^{(1)}}{\Psi^- (t)}\right |^2\, =\,\frac{1}{4}\, e^{-2 \Gamma\, t}. \nonumber 
\end{eqnarray}

\section{Oscillations of neutral pseudoscalar mesons. Functions
  $\textrm{F}_{N}$ and their properties}
\label{sec:DD}

In this Section we show the explicit form of the functions
$\textrm{F}_{N}$ which enter time-dependent inequality
(\ref{F_N}). Also we provide a Table~\ref{table:conditions} of correspondences between these
functions and sets of events for neutral pseudoscalar mesons that
violate (\ref{F_N}). 

\begin{eqnarray}
\label{FN-set}
\textrm{F}_{1} (x,\, r,\,\zeta,\,\lambda) &=&  
     \left | 
              g_+(t)\, -\, \frac{1}{2}\,\left ( \frac{q}{p} + \frac{p}{q}\right )\, g_-(t)
     \right |^2\, |g_+ (t)|^2\,\, e^{2 \Gamma\, t}; \nonumber \\
\textrm{F}_{2} (x,\, r,\,\zeta,\,\lambda) &=& 
     \left | 
              g_+(t)\, -\, \frac{1}{2}\,\left ( \frac{q}{p} + \frac{p}{q}\right )\, g_-(t)
     \right |^2\, e^{- \Delta \Gamma\,t /2}\,\, e^{\Gamma\, t};\nonumber \\
\textrm{F}_{3} (x,\, r,\,\zeta,\,\lambda) &=& 
      \left | 
              g_+(t)\, +\, \frac{1}{2}\,\left ( \frac{q}{p} + \frac{p}{q}\right )\, g_-(t)
     \right |^2\, |g_+ (t)|^2\,\, e^{2 \Gamma\, t}; \nonumber \\
\textrm{F}_{4} (x,\, r,\,\zeta,\,\lambda) &=& \left | 
              g_+(t)\, +\, \frac{1}{2}\,\left ( \frac{q}{p} + \frac{p}{q}\right )\, g_-(t)
     \right |^2\, e^{- \Delta \Gamma\,t /2}\,\, e^{\Gamma\, t}; \nonumber \\
\textrm{F}_{5} (x,\, r,\,\zeta,\,\lambda) &=& 
     \Big | g_+ (t) \Big |^2\,\, e^{- \Delta \Gamma\,t /2}\,\, e^{\Gamma\, t} ;\\
\textrm{F}_{6} (x,\, r,\,\zeta,\,\lambda) &=& 
     \Big | g_+ (t) \Big |^2\,\, e^{+ \Delta \Gamma\,t /2}\,\, e^{\Gamma\, t}; \nonumber \\
\textrm{F}_{7} (x,\, r,\,\zeta,\,\lambda) &=& 
     \left | 
              g_+(t)\, -\, \frac{1}{2}\,\left ( \frac{q}{p} + \frac{p}{q}\right )\, g_-(t)
     \right |^2\, e^{+ \Delta \Gamma\,t /2}\,\, e^{\Gamma\, t}; \nonumber \\
\textrm{F}_{8} (x,\, r,\,\zeta,\,\lambda) &=& 
     \left | 
              g_+(t)\, +\, \frac{1}{2}\,\left ( \frac{q}{p} + \frac{p}{q}\right )\, g_-(t)
     \right |^2\, e^{+ \Delta \Gamma\,t /2}\,\, e^{\Gamma\, t}. \nonumber 
\end{eqnarray}

\begin{table}[bt]
\caption{Table of correspondences between sets of events, functions
  $\textrm{F}_{N}$ related to these sets, and conditions of violation
  of the inequality (\protect\ref{F_N}). The events $\mathcal{S}_1
  (t_0)$ and $\mathcal{S}_2 (t)$ depend on the same ``directions'' of
  pseudoscalar mesons, so in the first column we provide only
  directions.  The event $\mathcal{S}_3 (t_0)$ is identical for all
  the sets of events and is not shown. 
\hfill\label{table:conditions}}
\bigskip
\begin{center}
\begin{tabular}{||c|c|l||}
\hline
\hline
 Set of events & Function &  Conditions of violations of the
                            inequality (\protect\ref{F_N})\\ 
\hline\hline 
$ \{M_1^{(2)}, M^{(1)}\}$ & $\textrm{F}_{1}$ & Violates for $B_s$-mesons \\
\hline
$\{M^{(2)}_1, \bar M^{(1)}\}$ & $\textrm{F}_{1}$ & Violates for $B_s$-mesons \\
\hline
$ \{M_1^{(2)}, M^{(1)}_H\}$ & $\textrm{F}_{2}$ & Violates for $B_s$-mesons \\
\hline
$ \{M_2^{(2)}, M^{(1)}\}$ & $\textrm{F}_{3}$ & Violates for $K$- and $D$-mesons \\
\hline
$ \{M_2^{(2)}, \bar M^{(1)}\}$ & $\textrm{F}_{3}$ & Violates for $K$- and $D$-mesons \\
\hline
$ \{M_2^{(2)}, M^{(1)}_H\}$ & $\textrm{F}_{4}$ & Violates for $K$- and $D$-mesons \\
\hline
$ \{M^{(2)}, M^{(1)}_H\}$ & $\textrm{F}_{5}$ & Violates for $K$-, $D$- and $B_s$-mesons \\
\hline
$ \{\bar M^{(2)}, M^{(1)}_H\}$ & $\textrm{F}_{5}$ & Violates for $K$-, $D$- и $B_s$-mesons \\
\hline
$ \{M^{(2)}, M^{(1)}_L\}$ & $\textrm{F}_{6}$ & Never violates \\
\hline
$ \{\bar M^{(2)}, M^{(1)}_L\}$ & $\textrm{F}_{6}$ & Never violates \\
\hline
$ \{M^{(2)}_1, M^{(1)}_L\}$ & $\textrm{F}_{7}$ & Never violates \\
\hline
$ \{M^{(2)}_2, M^{(1)}_L\}$ & $\textrm{F}_{8}$ & Never violates \\
\hline
\hline
\end{tabular}
\end{center} 
\end{table}

\end{document}